\documentclass[sigconf]{acmart}

\usepackage{subcaption}

\AtBeginDocument{%
  \providecommand\BibTeX{{%
    \normalfont B\kern-0.5em{\scshape i\kern-0.25em b}\kern-0.8em\TeX}}}

\copyrightyear{2023}
\acmYear{2023}
\acmConference[GECCO '23]{Genetic and Evolutionary Computation Conference}{July 15--19, 2023}{Lisbon, Portugal}
\acmBooktitle{Genetic and Evolutionary Computation Conference (GECCO '23), July 15--19, 2023, Lisbon, Portugal}
\acmDOI{10.1145/3583131.3590443}
\newcommand{\textred}[1]{#1}

\begin{document}

\title{Leveraging Human Feedback to Evolve and Discover Novel Emergent Behaviors in Robot Swarms}

\author{Connor Mattson}
\email{c.mattson@utah.edu}
\affiliation{%
  \institution{University of Utah}
  \city{Salt Lake City}
  \state{Utah}
  \country{USA}
}

\author{Daniel S. Brown}
\email{daniel.s.brown@utah.edu}
\affiliation{%
  \institution{University of Utah}
  \city{Salt Lake City}
  \state{Utah}
  \country{USA}
}

\renewcommand{\shortauthors}{Mattson and Brown}

%%
%% The abstract is a short summary of the work to be presented in the article.

%% Note: 200 words max.
\begin{abstract}
    Robot swarms often exhibit emergent behaviors that are fascinating to observe; however, it is often difficult to predict what swarm behaviors can emerge under a given set of agent capabilities. We seek to efficiently leverage human input to automatically discover a taxonomy of collective behaviors that can emerge from a particular multi-agent system, without requiring the human to know beforehand what behaviors are interesting or even possible. Our proposed approach adapts to user preferences by learning a similarity space over swarm collective behaviors using self-supervised learning and human-in-the-loop queries. We combine our learned similarity metric with novelty search and clustering to explore and categorize the space of possible swarm behaviors. We also propose several general-purpose heuristics that improve the efficiency of our novelty search by prioritizing robot controllers that are likely to lead to interesting emergent behaviors. We test our approach in simulation on two robot capability models and show that our methods consistently discover a richer set of emergent behaviors than prior work. Code, videos, and datasets are available at \url{https://sites.google.com/view/evolving-novel-swarms}.
\end{abstract}





\maketitle

% TODO: Make this graphic horizontal to use the best space.
\begin{figure*}
     \centering
     \includegraphics[width=\linewidth]{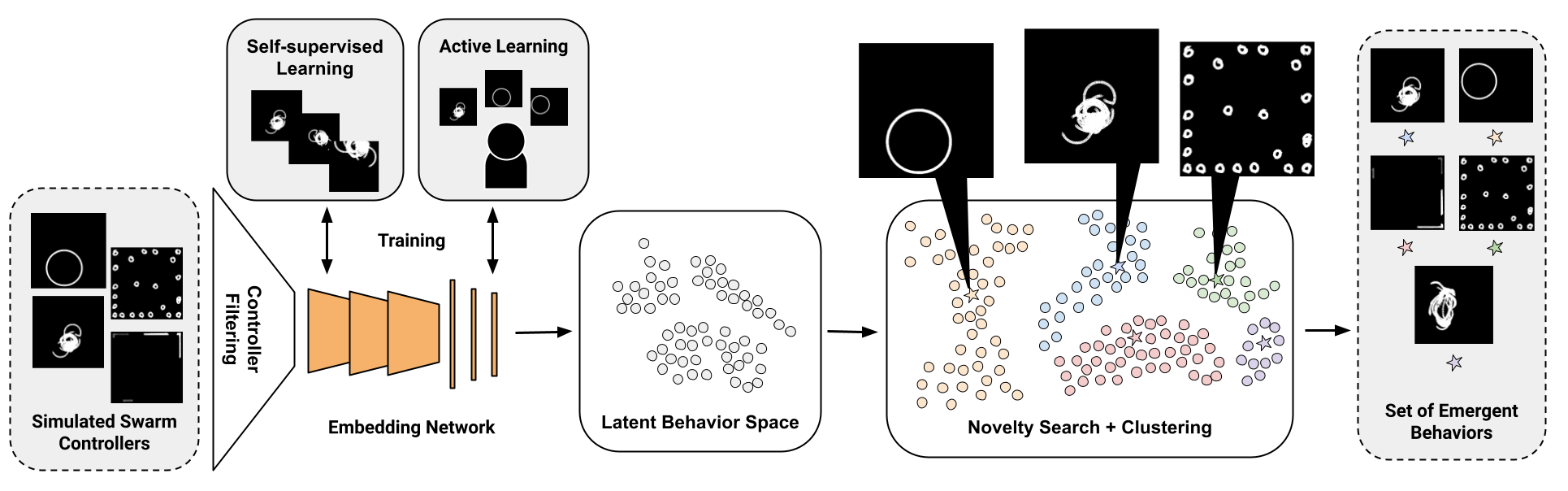}
     \caption{Self-Supervised learning and human-in-the-loop training to learn low-dimensional feature representations of complex swarm behaviors as the behavior space for exploring novel swarms. From left to right: Simulated swarm controllers are selectively sampled from using a filtering heuristic. A convolutional neural network trained by self-supervised contrastive learning over the controller space and active human-in-the-loop similarity queries. The learned embedding space is used to search for novel emergent behaviors which are then clustered to form a taxonomy of possible emergent behaviors.}
     \label{fig:teaser}
     \Description{A pipeline visual that shows the entirety of our approach as explained in the caption.}
\end{figure*}

\section{Introduction}
A robot swarm is a multi-robot system comprised of many independent agents. Swarm systems are decentralized, requiring agents to operate using only local knowledge of their surroundings, and often contain many small robots that have limited capabilities \cite{bayindir2007review,dias2021swarm}. By limiting a robot's ability to interact with and sense its environment, a single robot will not be able to produce complex behaviors alone. However, when several limited agents are allowed to operate in the same environment, the combined abilities of many robots may form a \textit{collective behavior}, an emerging pattern resulting from interactions between multiple agents and their environment.  

The synthesis of collective behaviors is a longstanding research pursuit. In evolutionary robotics~\cite{cliff1993explorations, nolfi2000evolutionary}, engineering these behaviors is approached by evolving swarm controllers that optimize a desirable objective. Prior work uses evolution to find controllers for aggregation~\cite{gauci2014evolving, trianni2003evolving}, shepherding~\cite{ozdemir2017shepherding}, object clustering~\cite{gauci2014clustering}, coverage~\cite{ozdemir2019spatial}, foraging~\cite{johnson2016evolving}, \textred{formation design~\cite{StolfiAutonomous, sperati2011self, st2018circle}, and collision optimization~\cite{MirhosseiniAdaptive}}. Given a fitness function that represents a swarm's performance with respect to a specific task, evolutionary search can be used to optimize the swarm's collective behavior.

However, swarm interactions are inherently complex.
%---given a desired swarm behavior, optimal multi-agent swarm control requires solving a difficult decentralized partially observable Markov decision processes (Dec-POMDP) \cite{huttenrauch2019deep}. 
The study of emergent behaviors has shown that even simple swarms may have unpredictable behaviors~\cite{parrish1999complexity,brown2018discovery,dorigo2021swarm}, and the complex global patterns that result from simple interactions are often difficult, if not impossible, to predict given initial conditions~\cite{adamatzky2010game}. However, as scientists and engineers, we want to understand, explore, and classify the types of emergent behavior that are possible. 
% , especially as we increase agent processing power, augment robot actuation, and add additional agents, and vary environments we want to understand, explore, and classify the types of behavior a given swarm is capable of.

In this paper, we seek to solve the following problem: \textbf{How can we leverage human-feedback to discover 
the set of possible emergent behaviors in a swarm of robots possessing a specific set of individual capabilities?} 
Our approach,
\textred{shown in Figure \ref{fig:teaser}}, leverages human similarity queries and contrastive deep learning to learn a latent behavior space that is useful for emergent behavior discovery.
Previous work evolves novel behaviors over a hand-crafted behavior space~\cite{brown2018discovery}, which requires domain expertise and a full specification of important behavioral features. Other evolutionary efforts to discover swarm behaviors include Swarm Chemistry~\cite{sayama2009swarm}, where a human acts as the fitness function in evolutionary search. In these approaches, the human selectively directs behavior exploration which limits the amount of exploration possible, due to the time and effort constraints of the human. We seek a solution to behavior discovery that does not require expert hand-crafted metrics, \textred{does not require the human to explicitly guide} behavior evolution, and does not require the human to know which behaviors are possible beforehand.

In the following sections, we present the following contributions:
% \begin{enumerate}
    % \item 
    (1) A novel approach to exploring and discovering emergent behaviors via a combination of human similarity queries and self-supervised contrastive learning to eliminate the need for hand-engineering behavior representations.
    % \item 
    (2) A set of heuristics that improve the efficiency of swarm novelty search by predicting ahead of time whether a controller will result in an interesting behavior.
    % \item 
    (3) We offer the first study of the emergent behaviors possible with a two-sensor computation free swarm and report two previously unknown emergent behaviors.  
    % \item 
    (4) We demonstrate the potential for our approach to generalize across different capability models. When tested on a single-sensor computation-free robot swarm, our approach discovers, on average, 3\% more behaviors than the prior state-of-the-art approach~\cite{brown2018discovery} which was hand-engineered for this specific capability model. When tested on a novel two-sensor computation-free capability model, our approach produces a 19\% increase in the number of emergent behaviors discovered.
% \end{enumerate}
% \vspace{-0.5cm}

\section{Problem Statement}\label{sec:problem}
All robots have limitations in their ability to actuate, sense, and compute information. Within a short period of time, a robot will only be able to move $X$ meters, collect information about its surroundings with $Y\%$ confidence, and compute it's next actions with $Z$ bytes of memory. We refer to a robot's \textit{capability model} as a set of sensing, actuation, and computation abilities used by the robot during its operation.

The desired output of a swarm of robots is often an emergent behavior. Emergent behaviors are global patterns that are the result of \textred{local} interactions between many agents \textred{within a particular environment}. 
% Like any robot, the abilities of a swarm can be amplified or hindered by the configuration of the environment that the robot is operating in. Therefore, it is also important to consider the impact of the environment on swarm behavior when considering behavior discovery.
While most swarm engineering seeks to optimize a swarm to excel at a particular task or behavior, we seek to uncover the complete taxonomy of emergent behaviors that could be exhibited by a swarm. Our research seeks to answer the following problem, first proposed by Brown et al.~\cite{brown2018discovery}.

\textbf{Behavior Emergence Problem:} \textred{We wish to find the set of possible emergent behaviors in a swarm of robots
possessing a specific set of individual capabilities.} To formalize this problem, let $\mathcal{C}$ represent the robot capability model of a swarm agent represented by the three-tuple $\langle S,M,A \rangle$ comprising a set of sensors $S$, available memory and computational processing resources $M$, and actuators $A$. Given an environment $\mathcal{E}$ and a controller space, $U(\mathcal{C})$, that is instantiated using the capabilities $\mathcal{C}=\langle S,M,A \rangle$, let $\Phi$ represent the mapping
% \begin{equation}
$
    \Phi: U(\mathcal{C}) \times \mathcal{E} \mapsto \mathcal{B},
    $
% \end{equation}
\textred{where $\mathcal{B}$ is the set of possible emergent behaviors. Given this formalism, our goal is to find} the image of the function $\Phi$. Specifically, what is the complete set of collective behaviors that emerges from the local interactions defined by $U(\mathcal{C})$ and $\mathcal{E}$?

In this paper, we focus on \textit{homogeneous swarms}, where all agents have the same capabilities and run the same controller. We seek to explore the behavior space, $\mathcal{B}$, by efficiently searching and evolving sets of controllers that result in a diverse taxonomy of emergent behaviors. In contrast to prior work by Brown et al. \cite{brown2018discovery}, we seek a solution to the behavior emergence problem that removes the need for a human to hand-engineer behavior features for each specific capability model of interest.

\section{Background and Related Work}
We begin by summarizing prior work and the necessary background needed to understand our methodology.

% \subsection{Human-Aligned Swarm Robotics}
% Other works in evolutionary swarm robotics successfully utilize human input at runtime to draw distinctions between desirable behaviors. Swarm Chemistry \cite{sayama2009swarm} performs interactive evolution to allow a human "alchemist" to participate in the selection, mutation, and crossover of behaviors and reported several examples of appealing emergent swarm behaviors.

\subsection{Novelty Search}
Evolutionary search excels at discovering high performing parameters with respect to a fitness function. This function often conveys success at a given task and allows for the search of an optimized solution, such as minimizing the distance from the robot to a goal or maximizing an agent's lifespan. Novelty Search \cite{lehman2011abandoning} uses novelty as the fitness metric, rewarding solutions whose behavior is novel when compared to previously found solutions. Formally, the novelty fitness function is defined as 

\begin{equation} \label{eq:novelty}
Novelty(b) = \frac{1}{k} \sum^{k}_{i=0} dist(b, \mathbb{A}_{i}),
\end{equation}
where $b$ is a behavior vector that lies in the space $\mathbb{R}^n$ and $\mathbb{A}_i$ is the $i$th-nearest neighbor of $b$ in the archive of all previously sampled behaviors, $\mathbb{A}$. By optimizing for divergence from known behaviors, the notion of what makes a behavior novel changes as the distribution of behaviors in the known set shifts. Over many generations, the archive of behaviors will reflect a diverse set of behaviors.  

In this paper we seek to learn a mapping from an emergent behavior down into its behavior vector, $b$. Prior work has looked at how to adapt a pre-specified behavior mapping for novelty search~\cite{gomes2014systematic,meyerson2016learning}. Other work has tried to hand-engineer generic behavior mappings that are task agnostic~\cite{gomes2013generic}; however, these approaches have been shown to provide the most benefit when combined with a specific objective-based fitness function~\cite{mouret2012encouraging}. Other work has used pretrained classification networks~\cite{nguyen2015innovation} or unsupervised auto-encoders to map from high-dimensional spaces to a latent behavior space~\cite{liapis2021transforming,grillotti2022unsupervised}. However, these approaches are designed to work with creative image or art generation or high-capability single agent systems. By contrast, our work seeks to explore and categorize the space of emergent behaviors in a swarm of robots. Similar to prior work, we learn a behavior embedding space using deep learning; however, rather than using auto-encoders, we use modern contrastive learning techniques and also leverage small amounts of human similarity labels to refine our pretrained latent space and better capture novel swarm behaviors.

% \subsection{Evolving over a Latent Embedding}

% \subsection{Human-Centered Swarm Learning}

\subsection{Swarm Behavior Search using Hand-Engineered Behavior Mappings}
We seek to discover a taxonomy of possible emergent behaviors in a swarm. Diversity has been shown to be beneficial in evolutionary robotics~\cite{mouret2012encouraging} and prior work has investigated searching for diverse ways to accomplish a specific task~\cite{engebraaten2018evolving}, including sequencing multiple different behaviors using a hierarchical controller~\cite{gomes2018approach}. However, most prior work focuses on using diversity to solve a particular problem, rather than seeking to discover and understand the full space of possible emergent behaviors.

Our work improves upon prior work on discovering emergent behaviors proposed by Brown et al. \cite{brown2018discovery}. Brown et al. explore the space of emergent behaviors that are possible given the computation-free capability model proposed by Gauci et al.~\cite{gauci2014evolving} (Fig. \ref{fig:brown capability model}) consisting of a single forward-facing line-of-sight sensor that can only relay a binary signal, 0 or 1, when another agent is detected~\cite{gauci2014evolving}. This single bit of information drives the velocities to the left and right wheels of a differential drive robot. Brown et al. represent a controller for the single-sensor model, $C_{1} \in U(\mathcal{C})$, as a vector of 4 values.
\begin{equation}
    C_{1} = [v_{l0}, v_{r0}, v_{l1}, v_{r1}],
    \label{eq: baseline controller}
\end{equation}
where $v_{\{l, r\}0}$ represents the velocities executed by the wheels when the sensor does not detect another agent and $v_{\{l, r\}1}$ represents the wheel velocities when another agent is detected. By representing agent instructions as a vector of real-valued numbers, controllers can be easily sampled from the continuous space $[-1, 1]^4$. 

\begin{table}
  \centering
  % \begin{small}
  \begin{tabular}{|l|l|}
    \hline
    \textbf{Feature}    & \textbf{Equation} \\ \hline
    Average Speed    & \(\frac{1}{N} \sum\limits_{i=1}^{N} \rVert v_i \lVert\) \\
    Angular Momentum & \(\frac{1}{R \cdot N} \sum\limits_{i=1}^{N} (v_i \times (x_i - \mu))\) \\
    Radial Variance  & \(\frac{1}{R^2 \cdot N} \sum\limits_{i=1}^{N} (\lVert x_i - \mu \rVert - \frac{1}{N} \sum\limits_{i=1}^{N}\lVert x_i - \mu \rVert )^2\) \\
    Scatter          & \(\frac{1}{R^2 \cdot N} \sum\limits_{i=1}^{N} \lVert x_i - \mu \rVert^2\)                \\
    Group Rotation   & \(\frac{1}{R \cdot N} \sum\limits_{i=1}^{N} (v_i \times \frac{x_i - \mu}{\lVert x_i - \mu \rVert})\)                \\ \hline
    \end{tabular}
    % \end{small}
 \caption{Hand-crafted behavior mapping proposed by Brown et al.~\cite{brown2018discovery}. These features were chosen by domain experts and tailored to a specific capability model. Rather than requiring hand-engineered features, we learn behavior features via self-supervised and human-in-the-loop learning.}
 \label{table:brown hand metrics}
\end{table}

Using an initial population of controllers, Brown et al. use novelty search to evolve over and explore the behavior space induced by controllers of the form given in Eq.~\eqref{eq: baseline controller}. Defining this behavior space requires the creation of a sufficient mapping that translates agent-environment interactions into low-dimensional behavior representations. Brown et al. introduce the behavior mapping in Table \ref{table:brown hand metrics}: a set of 5 metrics that map a collective behavior to a behavior vector in $\mathbb{R}^5$ and perform novelty search over these vectors. They then cluster the behavior \textred{vectors} in the archive and discover a taxonomy of 6 distinct emergent behaviors, one of which (milling) was previously unknown to emerge from computation-free robots. However, this approach relies on the carefully hand-crafted behavior feature space shown in Table~\ref{table:brown hand metrics}. Deriving this behavior mapping requires both an intuition about the behavior domain and some expertise in robot motion, both of which are difficult to obtain in the context of more complex capability models. 
% We seek to remove this requirement and shows that the results of Brown et al. are obtainable without human foreknowledge or expertise related to unknown swarm behaviors.

Rather than hand-engineering the behavior space, in this paper, we study to what extent this mapping can be learned. We leverage human-input and contrastive visual representation learning to develop a method for swarm behavior discovery that does not require domain expertise nor hand-crafted behavior mappings.

\subsection{Contrastive Visual Representation Learning}
Embedding images into a low-dimensional latent space through contrastive learning has emerged as a powerful, self-supervised way to learn features~\cite{hadsell2006dimensionality, chen2020simple, jaiswal2020survey}. Contrastive learning seeks to learn a similarity space where inputs belonging to the same group or class are clustered close together in $N$-dimensional space. 
% Specifically, self-supervised methods can learn feature similarity by performing small transformations to images to teach networks about similar features between different images. 
The triplet loss~\cite{balntas2016learning} facilitates this form of learning by sampling an anchor, $a$, and a dissimilar negative example, $n$, from a dataset. Then a transformation is applied to the anchor to produce a positive sample that is similar to a, $p = a'$. Formally, the triplet loss $\lambda$ is represented as 
\begin{equation}
    \label{eq: Triplet Loss}
    \lambda(a, p, n) = \text{max}\{||a - p||_2 - ||a - n||_2 + m, 0\},
\end{equation}
where \textred{the margin, $m$, represents a predefined target distance that the embedding will attempt to maintain between latent classes}. Triplet learning creates an embedding where the anchor and positive samples are closer to each other in $N$-dimensional space than the anchor and the negative. 
% In particular, triplet learning has been shown to generalize better and leads to faster convergence when compared to pairwise learning \cite{balntas2016learning}.

Triplet losses are often used to learn similarity models that capture how humans perceive images~\cite{agarwal2007generalized,tamuz2011adaptively,mcfee2011learning,cagatay2014kernels,ehsan2015kernel,hoffer2015deep,balntas2016learning}; however, to the best of our knowledge, we are the first to use a triplet loss to learn a similarity model of how humans perceive the emergent behaviors of robot swarms and to use this similarity metric to assist a human in discovering novel emergent behaviors.

\section{Approach}
As noted in Section~\ref{sec:problem}, our goal is to discover the set of emergent behaviors $\mathcal{B}$ that are possible given a swarm of robots with capabilities $\mathcal{C}$.
%learn a map $\Phi: U(\mathcal{C}) \times \mathcal{E} \mapsto \mathcal{B}$ that sufficiently projects similar behaviors into distinct areas of n-dimensional behavior space. 
We leverage self-supervised pretraining, human-in-the-loop swarm similarity learning, and heuristic filters to extend and generalize prior work on novel behavior search by Brown et al.~\cite{brown2018discovery}. We use self-supervised learning and active human similarity queries to replace the manually constructed behavior space defined in Table \ref{table:brown hand metrics} with a learned behavior latent embedding. Our approach promotes the discovery of emergent behaviors based on visual behavior alone, without reliance on hand-engineered representations of the emergent behavior space (Fig.~\ref{fig:teaser}). 
% We also contribute a set of regulations imposed on the controller space to reduce the number of trivial behaviors that are explored during evolution. 

In the following subsections, we introduce a behavior mapping based on the training of a deep neural network model through heuristic-based data filtering and two training periods. The first training period is a self-supervised triplet similarity training cycle. The second training period is an active learning approach aimed at aligning the pretrained model with a human's notion of swarm behavior distinctions. 
% Combined, we have a model that learns an embedding that replaces Brown et al.'s hand-crafted evaluation of behaviors prior to novelty search and clustering .  

\begin{figure*}
     \centering
     \begin{subfigure}[b]{0.16\linewidth}
         \centering
         \includegraphics[width=\linewidth]{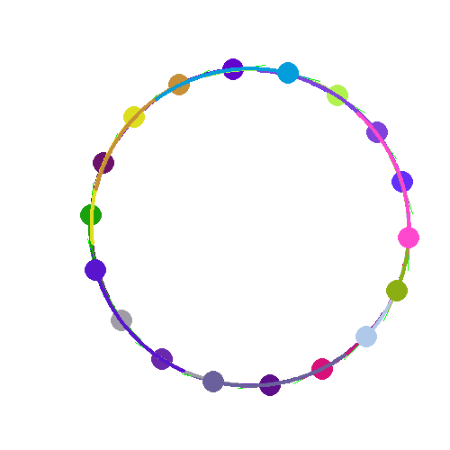}
         \includegraphics[width=\linewidth]{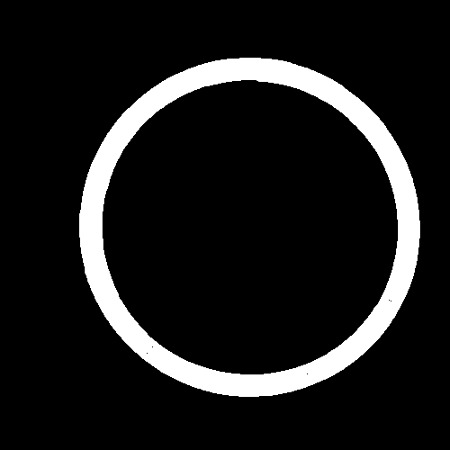}
         \caption{Cyclic Pursuit}
         \label{fig:visual-cyclic}
     \end{subfigure}
     \hfill
     \begin{subfigure}[b]{0.16\linewidth}
         \centering
         \includegraphics[width=\linewidth]{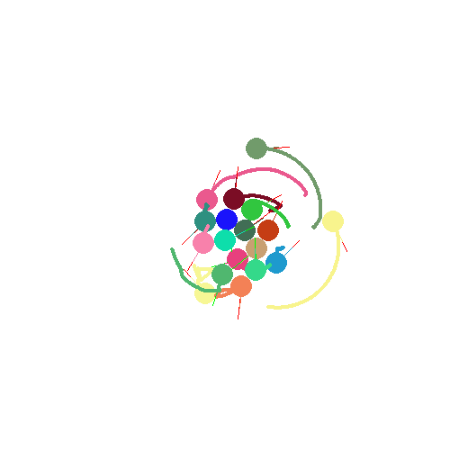}
         \includegraphics[width=\linewidth]{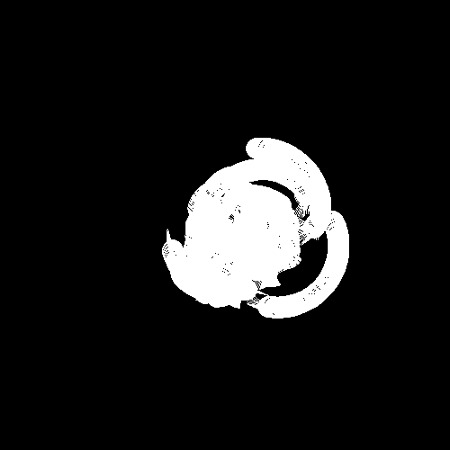}
         \caption{Aggregation}
         \label{fig:visual-aggregation}
     \end{subfigure}
     \hfill
     \begin{subfigure}[b]{0.16\linewidth}
         \centering
         \includegraphics[width=\linewidth]{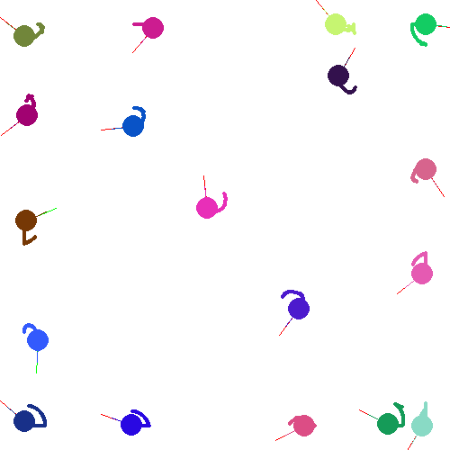}
         \includegraphics[width=\linewidth]{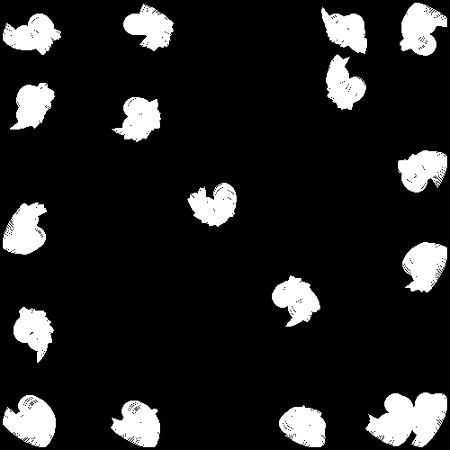}
         \caption{Dispersal}
         \label{fig:visual-dispersal}
     \end{subfigure}
     \hfill
     \begin{subfigure}[b]{0.16\linewidth}
         \centering
         \includegraphics[width=\linewidth]{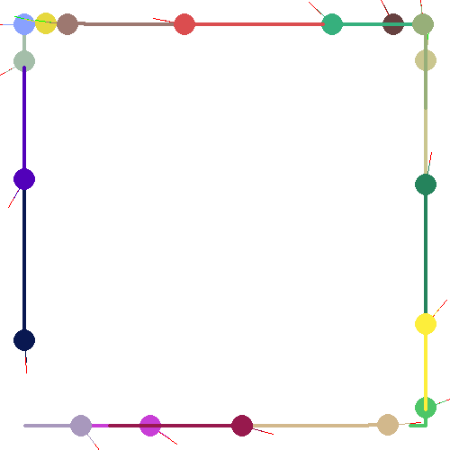}
         \includegraphics[width=\linewidth]{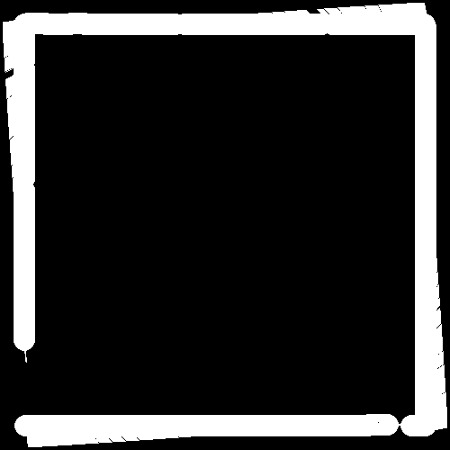}
         \caption{Wall Following}
         \label{fig:visual-wall-f}
     \end{subfigure}
     \hfill
     \begin{subfigure}[b]{0.16\linewidth}
         \centering
         \includegraphics[width=\linewidth]{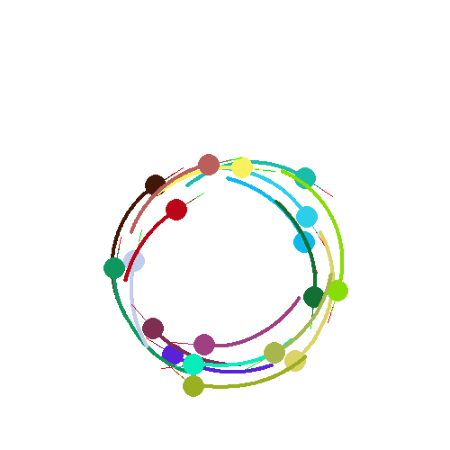}
         \includegraphics[width=\linewidth]{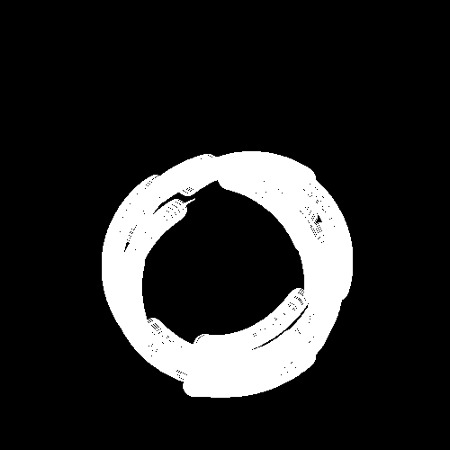}
         \caption{Milling}
         \label{fig:visual-milling}
     \end{subfigure}
     \hfill
     
    \caption{Swarm behaviors (top) and their corresponding grayscale visual representations (bottom). Behaviors are simulated over a fixed time horizon and then converted into 50x50 images of the agents' trajectory over the final frames of simulation.}
    \Description{On top, a row of 5 images with colored agents producing varying emergent behaviors. On bottom, the corresponding grayscale trajectories, where the trajectories of the agents are shown in white on top of a black background.}
    \label{fig:behaviors and visual representations}
\end{figure*}

\subsection{Data and Network Architecture}
We propose the use of a deep neural network that \textred{learns behavioral distinctions between swarm trajectories} and embeds them into a low-dimensional latent space. \textred{Following prior work by} Brown et al.~\cite{brown2018discovery}, the robot workspace is represented in 2D, where the features and patterns of swarms all exist on a plane. \textred{We investigate homogeneous swarms of differential drive agents, where patterns that emerge in prior work can be succinctly represented by collapsing the final steps of swarm agents into a single image (Fig. \ref{fig:behaviors and visual representations}).} We embed this image data into a latent vector using a Convolutional Neural Network that passes information through convolutional and fully connected layers. Our approach focuses on the 2D workspace \textred{and single-channel trajectory inputs}, but our methodology can be extended to other forms of workspaces, as long as the swarm's trajectories and collective behaviors can be represented visually over time.
We take advantage of two stages of training that allows \textred{our embedding network to learn} useful information about swarm \textred{behaviors}, despite having no labeled behavioral data and requiring no prior knowledge of the space of possible emergent behaviors. 

\subsection{Self-Supervised Pretraining}
We train our network to learn similarities between input features using an self-supervised approach. We utilize a triplet loss (Eq.~\ref{eq: Triplet Loss}), that allows us to teach a network the relational differences between behaviors that we present to it in an anchor, positive, and negative tuple. Following the techniques presented by Chen et al. \cite{chen2020simple}, we synthetically develop triplets by transforming a randomly sampled anchor into a positive example through a series of transformations. Chen et al. thoroughly explores the potential ways to synthetically create positive triplet samples, and reports that random cropping and color distortion work best as a pair of transformations that create synthetic anchors. We follow the same procedure recommended by Chen et al.~\cite{chen2020simple}. We obtain a triplet by first simulating 2 randomly sampled controllers from the controller space, which act as our anchor and our negative. Then we transform our anchor into our positive sample through random rotation and cropping transformations. This approach allows us to generate a large self-supervised dataset of triplets. We then perform batch training using Triplet Loss to create a network that has an initial understanding of swarm behaviors prior to the second stage of training. 

\subsection{Active Learning via Human Feedback}
We aim to maximize the accuracy of our learned behavior embedding, which is a daunting task for self-supervised pretraining alone. However, studies show that self-supervised training followed by supervised training with only 1\% of the truth labels can significantly improve network performance \cite{chen2020simple, zhai2019s4l}. Thus, we hypothesize that adding a human to the training loop will result in a more accurate learned similarity metric by enabling us to fine-tuned the embedding space based on human similarity queries.

We query a human to provide truth data for a small fraction of the behaviors simulated to form the training dataset. For each query, the human is presented with an image of the simulated behavior trajectory and asked to assign it to one of the known behavior classes, or as a new behavior belonging to an unknown class. By obtaining behavioral class labels, we can synthetically generate thousands of triplets from class labels alone by iterating over all pairs of elements that belong to the same class, which act as our anchor and positive sample, and creating triplets using negative samples from all members of other classes. Triplet generation of this nature is far more informative to our network than simply asking a human to provide true/false labels on triplets themselves, which would result in only one triplet for every three behaviors considered, compared with the creation of all possible triplets that can be generated from a list of known classes. 

\subsection{Heuristic Filtering}
\textred{Our self-supervised training approach relies on training on a diverse set of behaviors obtained by simulating randomly sampled swarm controllers. However, many controllers may not lead to interesting emergent behavior within the provided simulation time.} When considering the controller space spanned by $[-1, 1]^4$ (Eq. \ref{eq: baseline controller}), the small clusters of controllers corresponding to interesting emergent behaviors are sparse compared to the large number of controllers that result in uninteresting behaviors (Appendix \ref{Appendix: Informed Heuristics}). \textred{Simulating non-interesting controllers wastes computation time and these controllers can skew the data distribution away from interesting emergent behaviors,} leading to a poorly trained network~\cite{johnson2019survey}. 

We propose general-purpose heuristics to filter uninteresting controllers based on the following assumptions:
\begin{enumerate}
    \item Agents that operate with very low velocities will not converge upon an interesting collective behavior within the simulation horizon, $T$.
    \item Agents that fail to explore outside of their initial state will not converge upon interesting behavior.
    \item \textred{Emergent behavior is unlikely if agents do not change their individual behavior when they sense other agents.}
\end{enumerate}

Filtering is used as a preprocessing step to eliminate unnecessary exploration over controllers that can be safely assumed will not lead to emergent collective behaviors. A complete synthesis of these filters can be found in Appendix \ref{Appendix: Informed Heuristics}.

\subsection{Evolution and Clustering} 
After training a network that maps from visual swarm behaviors (Fig.~\ref{fig:behaviors and visual representations}) into a latent behavior space, we follow the methodology proposed by Brown et al.~\cite{brown2018discovery} to discover and report a taxonomy of behaviors using novelty search and $k$-medoids clustering. Using an initial sample of controllers, we evolve new controllers through mutation and crossover of the highest performing controllers from the previous generation where fitness is determined by the novelty metric (Eq.~\ref{eq:novelty}) \cite{lehman2011abandoning}. After evolution, we perform a $k$-medoids clustering over the entire set of explored behaviors in the archive $\mathbb{A}$. The $k$ behaviors corresponding to the medoids form our approximate taxonomy over the space of possible emergent behaviors. 

\begin{figure}[t]
     \centering
     \begin{subfigure}[b]{0.4\linewidth}
         \centering
         \includegraphics[width=\linewidth]{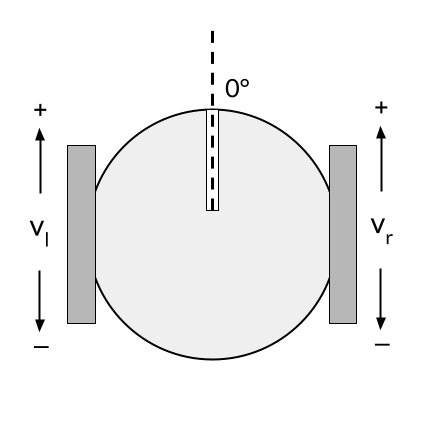}
         \caption{Single-Sensor}
         \label{fig:brown capability model}
         \Description{A differential drive agent with a circular body. The agent contains a single sensor that faces directly forward, parallel to the direction of the wheels.}
     \end{subfigure}
     \hspace{0.5cm}
     \begin{subfigure}[b]{0.4\linewidth}
         \centering
         \includegraphics[width=\linewidth]{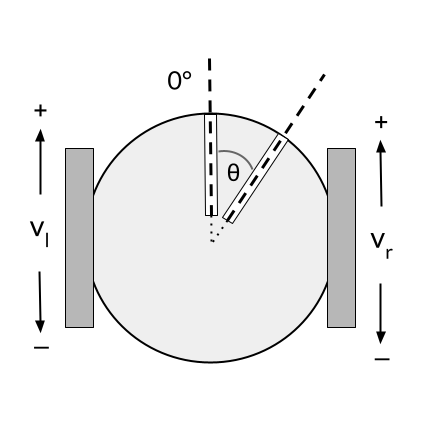}
         \caption{Two-Sensor}
         \label{fig:augmented model}
         \Description{A differential drive agent with a circular body. The agent contains two sensors: one that faces directly forward, parallel to the direction of the wheels and another sensor that is offset a some angle theta from the forward facing direction.}
     \end{subfigure}
     \caption{We explore emergent behaviors that result from a single- and two-sensor computation-free capability model, where each sensor is a binary line-of-sight sensor.}
    \label{fig:new capability models}
\end{figure}

\section{Experiments \& Results}
% Our approach proposes a combination of self-supervised pretraining, network refinement via active learning, and a filtering to improve the training data passed through our network. We describe two robot capability models that will act as our domains for experimentation. 
In this section, we quantify the potential of our methods by verifying that our learned latent embedding reflects the true feature similarities observed in a labeled validation dataset. We also determine the viability of using our embedding network as a replacement for a hand-engineered behavior mapping when performing emergent behavior exploration and categorization.

\subsection{Robot Capability Models}

% TODO 7: Revise Augmented Controller Model to match our new model
We consider the use of two different capability models. Both models utilize mobile robots that operate on differential drive kinematics where velocities are commanded to the left and right wheels separately. The first is the computation-free capability model proposed by Gauci et al.~\cite{gauci2014evolving} where agents contain a single line-of-sight binary sensor (Fig. \ref{fig:brown capability model}). The second is an augmented two-sensor capability model, where agents contain a forward-facing binary sensor, and an angled sensor oriented at some angle $\theta$ offset from the forward direction (Fig. \ref{fig:augmented model}). In the two-sensor model, the angle for the second sensor is homogeneous across all agents in the swarm, but may vary from population to population as our algorithm evolves to find sensor angles that result in interesting behaviors. Because our sensing ability has increased, the controller for the two-sensor capability model is adapted from Equation. \ref{eq: baseline controller} to form
\begin{equation}
    \label{eq: two sensor controller}
    C_{2} = [v_{l0}, v_{r0}, v_{l1}, v_{r1}, v_{l2}, v_{r2}, v_{l3}, v_{r3}, \theta],
\end{equation}
where $v_{\{l, r\}2}$ represents the velocities executed when the additional sensor is toggled on and $v_{\{l, r\}3}$ represents the velocities executed when both additional sensors are toggled on.

\subsection{Simulation and Data}
To run our experiments we built a light-weight and easily extensible Python simulator\footnote{Code available here: \url{https://sites.google.com/view/evolving-novel-swarms}} using Pygame~\cite{pygame.org}.
 For consistency, we follow Brown et al.'s experimentation and initialize our agents in a 500x500 units environment surrounded by rigid walls~\cite{brown2018discovery}. Friction between agents and the ground, walls, and other agents is ignored. Collisions between an agent and another agent or a wall results in the agent freely sliding against the object it is colliding with. Agent sensors detect other agents, but do not detect the environment walls. All agents have a body length of 10 units and infinite sensing distance.

We set a horizon on our simulation by limiting the amount of time that agents are allowed to interact within the environment. For each simulation, 24 agents are instantiated into the environment at random positions and orientations and execute the homogeneous controller $c \in U(\mathcal{C})$. Our experiments use a time horizon of $T$=1200.
%, after which the information is collected about the swarm's behavior over a window marking the end of simulation.
Controllers sampled from $U(\mathcal{C})$ are homogeneously applied to each agent in the swarm. The final 160 frames of simulation collapsed into a single-channel, 50x50-pixel, grayscale image (see Fig.~\ref{fig:behaviors and visual representations}). 
% This results in images that are entirely black apart from the trajectories of agents within the last 80 timesteps. 
% Images are resized to 50x50 pixels. 
% To convert this behavior into a image, we paint a trajectory of the last 80 frames into one single greyscale image and resize it to 50x50 as shown in Fig.~\ref{fig:behaviors and visual representations}. 

Our network takes a $B$x1x50x50 tensor as input, where $B$ is the size of the input batch. This tensor passes through 3 Conv2D layers and 3 fully-connected layers into an output vector, $y \in \mathbb{R}^N$. We use $N$=5 for a fair comparison with the dimensionality of Brown et al.'s hand-crafted behavior space~\cite{brown2018discovery}. 
% We use $n=9$ \textbf{TODO: CHeck number with connor} for the two-sensor model. 
The CNN trains using Triplet Learning (Eq.~\ref{eq: Triplet Loss}), the Adam Optimizer, a learning rate (LR) of 0.08, and a 1e-6 weight decay. We also impose a LR scheduler that reduces the LR after 15 epochs of loss plateau.

\textred{To obtain the training data, we simulate 10,000 randomly sampled controllers, converting each controller simulation into an image as described above}. For every epoch of self-supervised training, we randomly sample 16,384 $\langle$anchor, negative$\rangle$ pairs from this training data and augment each anchor to synthesize a positive sample, creating 16,384 $\langle$anchor, positive, negative$\rangle$ triplets. 
We transform the anchor into our positive element by randomly cropping the image to [0.6, 1.0] times its original width and height and then randomly rotating the crop by an angle, $\alpha \in \{90^o, 180^o, 270^o\}$. \textred{Although it is possible for anchor and negative samples to be similar in behavior, it is unlikely that the negative example is closer to the anchor than a simple augmentation of the anchor.} We use a large batch size of 4096 triplets, which has been shown to benefit contrastive learning by providing more negative examples to the model~\cite{chen2020simple}. We stop training at 500 epochs or when the average loss over the last 10 epochs is less than 1e-3.

% During HIL Learning, we use a synthetic human to answer queries, which has been provided access to the class labels for the self-supervised training data, allowing 

\begin{figure}
     \centering
     \includegraphics[width=0.7\linewidth]{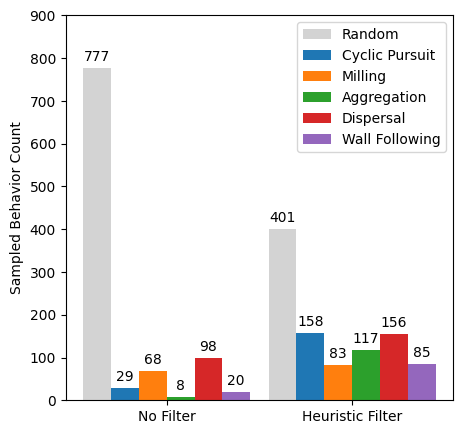}
         \caption{Class distributions for a uniform sampling of 1000 controllers from the controller space with and without our filtering heuristic.}
     \label{fig:filtering distribution}
     \Description{A bar plot comparing the distributions of samples without a controller filter (left) to the distribution of samples obtained with a controller filter (right). The no filter distribution shows 77 percent of sampled controllers are uninteresting and random, where only 40 percent of samples are random under our filter.}
\end{figure}

\begin{table}
    \centering
    % \begin{small}
    \begin{tabular}{lcc}
        \hline
         & \multicolumn{1}{l}{Single-Sensor} & \multicolumn{1}{l}{Two-Sensor} \\ \hline
        \textit{Baseline}                & \multicolumn{1}{l}{}              & \multicolumn{1}{l}{}           \\
        Random Network Initialization    & 69.13                             & 61.37                          \\
        Hand-Crafted (Brown et al.)      & 84.65                             & 67.52                          \\ \hline
        \textit{Latent Embedding (Ours)} & \multicolumn{1}{l}{}              & \multicolumn{1}{l}{}           \\
        Self-Supervised Pretraining         & 86.51                             & 72.11                          \\
        Pretraining + HIL + Filtering    & \textbf{88.09}                    & \textbf{74.33}                 \\ \hline
        \end{tabular}
        % \end{small}
    \caption{L2 embedding accuracy for two baseline embeddings and our approach on two robot models. Results were obtained on a validation dataset of 200 labeled swarm behaviors.}
    
    \label{table:Embedding Space Accuracy}
\end{table}

\begin{table*}
\small
    \centering
    % \begin{small}
    \begin{tabular}{lccccccccc}
    \hline
    & Cycp                 & Aggr                 & Disp                 & Mill                 & Wall-F                & N-Cycle              & C-Path               & Rand                 & Average Distinct Behaviors \\ \hline
    \textit{Single-Sensor Model (k=12)} & \multicolumn{1}{l}{} & \multicolumn{1}{l}{} & \multicolumn{1}{l}{} & \multicolumn{1}{l}{} & \multicolumn{1}{l}{} & \multicolumn{1}{l}{} & \multicolumn{1}{l}{} & \multicolumn{1}{l}{} & \multicolumn{1}{l}{}       \\
    Random Controller Sample                & 0.9                  & 0.1                  & 1.6                  & 0.6                  & 0.2                  & -                    & -                    & 8.6                  & 3.2                        \\
    Random Controller Sample + Filter       & 2.1                  & 1.9                  & 2.0                  & 0.7                  & 1.7                  & -                    & -                    & 3.6                  & 4.9                        \\
    Brown et al.                            & 1.4                 & 1.0                 & 2.0                 & 1.3                 & 0.9                 & -                    & -                    & 5.4                 & 5.8                       \\
    Brown et al. + Filter                   & 0.5                    & 1.0                    & 1.7                    & 2.8                    & 1.0                    & -                    & -                    & 5.0                    & 5.4                          \\
    HIL Latent Embedding (Ours)                 & 1.1                 & 1.0                 & 2.9                 & 1.4                 & 1.2                 & -                    & -                    & 4.4                 & \textbf{6.0}                       \\ \hline
    \textit{Two-Sensor Model (k=16)}               &                      &                      &                      &                      &                      &                      &                      &                      &                            \\
    Random Controller Sample                & 0.8                  & 0.6                  & 2.0                  & 0.8                  & 1.1                  & 0.2                  & 0.1                  & 7.2                  & 4.4                        \\
    Random Controller Sample + Filter       & 0.5                    & 1.2                    & 2.1                    & 1.2                    & 1.6                    & 0.2                    & 0.0                    & 9.2                    & 4.8                          \\
    Brown et al.                            & 0.3                    & 0.6                    & 1.9                    & 2.0                    & 1.3                    & 0.0                    & 0.3                    & 8.8                    & 5.1                       \\
    Brown et al. + Filter                   & 0.6                    & 2.3                    & 1.2                    & 1.7                    & 1.7                    & 0.1                    & 0.0                    & 8.2                    & 5.3                          \\
    HIL Latent Embedding (Ours)                 & 0.7                    & 1.0                    & 4.0                    & 1.3                    & 1.9                    & 0.6                    & 0.0                    & 6.5                    & \textbf{6.1}                       \\ \hline
    \end{tabular}
    % \end{small}
    \caption{Average number of swarm behaviors returned by random sampling, Brown et al.\cite{brown2018discovery}, and our Human-Trained Latent Embedding. Results are averaged over during the final 10 generations of evolution for two different robot models. Tracked behaviors are, from left to right, Cyclic Pursuit (Cycp), Aggregation, (Aggr), Dispersal (Disp), Milling (Mill), Wall Following (Wall-F), Nested Cycles (N-Cycle), Concave Path (C-Path), and Random (Rand).}
    
    \label{table:Behavior Emergence}
\end{table*}

\subsection{Heuristic Controller Filtering}
We validate the results of our filter on a random sample of controllers from the controller space. Fig.~\ref{fig:filtering distribution} shows the impact of our heuristic filter on the distributions of classes obtained by uniformly sampling 1000 controllers from the single-sensor controller space. We find that our heuristic provides an ample reduction in the number of uninteresting behaviors from 77.7\% without filtering to 40.1\% with filtering. We also find that previously underrepresented classes are sampled with higher frequency when our filter is applied. These results validate that our heuristic filtering removes uninteresting controllers and helps guide our controller sampling towards interesting emergent behaviors. We further examine the impact of this heuristic on the controller space in Appendix \ref{Appendix: Informed Heuristics}.

\subsection{Latent Representation Performance}
To validate the quality of our learned latent space, we measure the accuracy of our behavior embedding and compare results against two baseline latent spaces. For all experiments, we use a validation set of 200 labeled swarm behaviors to calculate the proximity of same-class behaviors in the embedded space. Data contained in the validation set is withheld from all training cycles and differs only between the baseline model and the two-sensor model. All results are averaged over 3 trials. 

We examine the quality of the behavior space representations using this validation data on two comparative baselines: Brown et al.'s~\cite{brown2018discovery} original behavior mapping and our deep network initialized with random weights and no training. To account for the variance in accuracy resulting from randomly initialized weights, results for Random Weight Initialization experiments are averaged over 30 trials. We measure the accuracy of our behavior embeddings across the set of all admissible triplet combinations in the labeled validation data, $\{$ $\langle a, p, n \rangle$ | $Class(a) = Class(p) \neq Class(n), a \neq p \neq n\}$. Accuracy measurements are gathered using Euclidean Distance (L2), where for an admissible triplet in the test set,
\begin{equation}
    L2(a, p, n) = \left\{
        \begin{array}{lr}
            Correct & \text{if } ||a - p||_2 < ||a - n||_2 \\
            Incorrect & \text{otherwise}
        \end{array}
    \right\}.
\end{equation}

Our results show similar findings across both robot capability models (Table \ref{table:Embedding Space Accuracy}). We find that a random embedding of behaviors performs significantly worse than both Brown et al's hand-crafted feature representations and our latent learning approach. We show that pretraining our network using only self-supervised contrastive learning is a powerful method that outperforms even the hand-crafted representation of the behaviors by 1.86\% on the single-sensor capability model and 4.59\% on the two-sensor model. Furthermore, we show that the best results are obtained when combining pretraining with human labeled data and our heuristic filter, improving upon the hand-crafted baseline by 3.44\% on the single-sensor robot model and 6.81\% for the two-sensor model.

% Table \ref{table:Embedding Space Accuracy} shows the quality of our pretraining approach alone and our pretraining in combination with HIL learning and Heuristic filtering. We find that for both capability models, our latent embedding approach (pretraining + human-in-the-loop + filtering) achieves an L2 accuracy improvement over both baselines. Furthermore, we validate the importance of human-led refinement and heuristic filtering by showing an improvement over pretraining.  

\subsection{Evolving New Behaviors}
Using the learned latent encoding, we use novelty search to evolve a full span of behaviors for the one- and two-sensor capability models. We evolve over our behavior space for 100 generations at 100 population each. For each experiment, we evaluate the average number of distinct emergent behaviors returned by $k$-medoids for the final 10 generations of evolution. This averaging over medoids helps capture the impact of the behavior space shifting.

We compare the diversity of behaviors obtained through our approach to a random sampling from the controller space, Brown et al.'s hand-crafted metrics and our Human-in-the-loop (HIL) latent embedding. In random sampling, a batch of $k$ uniformly sampled controllers is evaluated to determine the number of distinct behaviors in the batch and which behavior classes are represented. The average distinct behaviors and the average number of appearances from each behavior class across 10 batches are recorded. We find that the average number of distinct behaviors is improved when a filter is applied to the random sampling for both capability models.

Both the hand-crafted baseline and our proposed method confidently outperform a random selection of controllers. For the single-sensor model, we cluster the behavior space using $k$-medoids and k=12. When considering the single-sensor capability model, our method returns all 6 behaviors on average, slightly outperforming the hand-crafted metrics, which return 5.8 behaviors on average. Our method further verifies Brown et al.'s approach by uncovering the same 6 emergent behaviors for the single sensor model. 

We conjecture that a two-sensor robot model will have more interesting emergent behaviors than our single sensor behavior space, so we increase the value of k to 16 in order to examine more medoids for all experiments on our two-sensor model. Experiments on the two-sensor model show that our approach reports 19\% more unique behaviors on average when compared to Brown et al.'s hand-crafted behavior space. Our model discovers 6.1 distinct behaviors on average, over Brown et al.'s 5.1 behaviors, in a controller space that is exponentially larger than the single-sensor controllers. We also find a small improvement in the average number of distinct behaviors returned when combining Brown et. al's approach with a filtered behavior space, further reinforcing the benefits of filtering controllers even when applied to other methods.

\begin{figure}[h]
     \centering
     \begin{subfigure}[b]{0.4\linewidth}
         \centering
         \includegraphics[width=\linewidth]{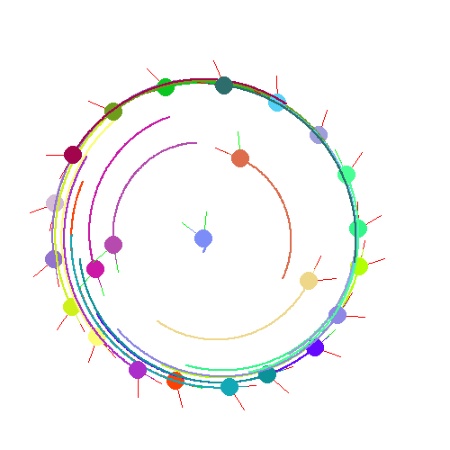}
         \caption{Nested Cycle}
         \label{fig:nested cycles}
         \Description{A colorful swarm trajectory showing a ring of outer agents surrounding a nested ring of agents that are milling in the center.}
     \end{subfigure}
     \hspace{0.5cm}
     \begin{subfigure}[b]{0.4\linewidth}
         \centering
         \includegraphics[width=\linewidth]{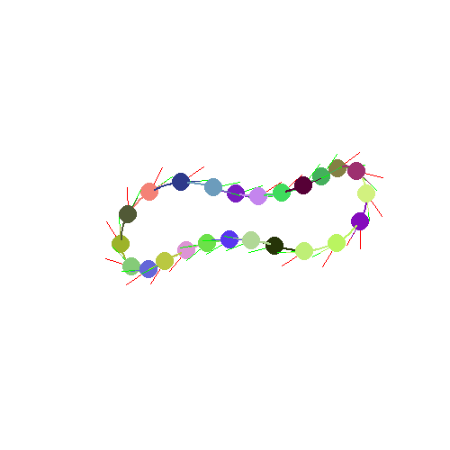}
         \caption{Concave Path}
         \label{fig:concave path}
         \Description{A colorful swarm trajectory showing a cyclic path that has concavity, biasing the path closer to the centroid of the swarm at some points along the trajectory.}
     \end{subfigure}
     \caption{Two new emergent behaviors discovered during experimentation for a computation-free two-sensor capability model. (a) A nested cycle, where a perimeter of agents surrounds an inner ring of cycling agents and (b) A concave path, where wave-like cyclic patterns result in a path where agents do not form a convex shape. }
    \label{fig:newly discovered behaviors}
\end{figure}

Our experiments produce two previously unknown swarm behaviors for a computation-free, two-sensor capability model. The first is a nested cycle (Fig.~\ref{fig:nested cycles}), where the swarm agents form two layers of cyclic pursuit. This behavior appears to be a hybrid of milling and cyclic pursuit, where exterior agents lock the inner agents into the center of the circle and interior agents mill without disturbing the outer circle's activity. This kind of behavior could only be facilitated using two sensors at this specific angle---the exterior agents utilize only their forward facing sensors,  while the interior agents utilize both of their sensors, resulting in the smaller radius of the inner circle. 

The second new emergent behavior is a concave path (Fig.~\ref{fig:concave path}), where a complete cycle is formed that preserves a structure that is neither circular nor elliptical. This behavior places its second sensor slightly offset from the first. When the forward-facing sensor is activated, an agent turns inward, but stops and turns outward when the second sensor is also activated. This switching between turning slightly inward and slightly outward while still driving forward results in a cycle that contains concavity along the path's circumference. Sampling this behavior is rare: random sampling only obtains this behavior once out of 160 samples and our approach does not discover this behavior. However, despite missing this behavior in our approach, our method succeeds in consistently finding more distinct behaviors on average than other approaches.   

% The two sensor model is naturally an extension of the single sensor space. 
\textred{Note that all behaviors possible with a single-sensor are possible with two sensors:} if you consider a single sensor controller $[a, b, c, d]$, the equivalent controller in the two sensor configuration space is $[a, b, c, d, a, b, c, d, \theta]$, which ignores the additional angled sensor and instead operates only with regards to the original forward-facing sensor. Because of this, we know that all 6 behaviors from the single sensor model can be discovered in the higher dimensional space. The results for our latent exploration reflect this, showing that all 6 emergent behaviors from the one-sensor capability model are also discovered for the two-sensor capability model. Videos of these behaviors can be found on our website.

\section{Discussion and Future Work}
% It is worth noting that both a hand crafted approach and our learning approach suffer from the curse of dimensionality. Specifically, consider the space spanned by the two controller models we experimented on. Earlier, we mention that the the size of our discretized single-sensor controller space is $21^4=194,481$ controllers, and our discretized two-sensor controller space is $21^8 \times 10 \approx 3.77x10^{12}$ controllers.

Because our model relies on pretraining to learn a latent embedding, our method relies on the assumption that the random samples of controllers that are used for pretraining and for human in the loop labeling reflect the true diversity of the behavior space. As the size of our controller space grows, it becomes more difficult to train a neural network to capture all the necessary behavior features through random sampling alone. Interesting areas of future work include 
% examining to what extent we can leverage symmetry or use compression to reduce the size of the controller space when searching for novel behaviors and developing more efficient sampling approaches that promote variance and diversity in sampled batches. In particular, we are excited about 
the possibility of applying and extending techniques from quality-diversity search~\cite{pugh2016quality,gravina2018quality,fontaine2021differentiable} and Markov chain Monte Carlo sampling~\cite{neal2011mcmc,hoffman2014no} to more efficiently search for emergent behaviors.  

Our experiments study a two-sensor computation-free robot capability model. To the best of our knowledge, we are the first to explore the space of behaviors that can be achieved from this capability model and our results have uncovered two previously unknown emergent behaviors that can result from computation-free swarms, along with the six behaviors that can result with either a one- or two-sensor capability model. We acknowledge that a more thorough exploration of the two sensor controller space is required to confidently determine that there are only eight emergent behaviors that exist for this capability model. Our results indicate that one of the behaviors, Concave Path, is quite rare and difficult to find with any of the behavior search methods we evaluate and is \textit{not} found using our proposed method. It is likely that there are other interesting emergent behaviors that current methods do not capture. Developing improvements to our method's pretraining, evolution, human query efficiency, and learned latent behavior mapping all have the possibility to reveal additional novel behaviors and facilitate a better understanding of the relationship between individual capabilities and the global emergent behaviors.

Given diverse input data, we show that we can train a latent behavior space by mapping a behavior to a single 50x50 greyscale trajectory image which is then embedded using our trained neural network. However, mapping the behaviors into an image will not always be possible or natural for all capability models. 
It is possible that behavior features on complicated models may require multi-frame video or point cloud representations. 
Behaviors that are defined, in part, by non-visual elements such as wireless communication would only have a good latent representation if the non-visual data is also represented in the behavior mapping. We believe that recent advances in embedding multiple modalities such as visual observations, sound, and language into common latent spaces~\cite{al2022zero} provides a promising way to encode more than just the observed motion patterns of a swarm into a behavior vector for use in novelty search and emergent behavior discovery. \textred{Future work should also focus on behavior discovery that can more readily be applied in real-world robotics by incorporating noise and variance into the actuation and sensing capabilities of agents.}

An important aspect to consider in human-in-the-loop learning is the investment of the human's labeling efforts versus the return on the network's accuracy gain. We show that our methods allow the human to refine the latent space by offering distinctions between behaviors that cannot be fully learned through a pretraining process alone. However, it is likely that the human has features or patterns in mind that they think are interesting, even if they do not initially know what types of emergent behaviors are possible. An interesting direction for future work is to explore efficient methods that allow a human to guide the search for novel emergent behaviors without requiring lengthy sessions of query labeling. 

% We show that evolutionary novelty search offers a sufficient exploration of the range of behaviors possible for a given capability model. 
% Clustering as a representation of behavior taxonomy works well in the context of our very simple agents. 
% However, using a $k$-medoids clustering may pose problems for robots of additional complexity where large behavior spaces result in small pockets of interesting behaviors becoming clustered with uninteresting behaviors. Furthermore, selecting an appropriate number of clusters, $k$, is difficult when there is no well-defined understanding of the range of behaviors that can be explored in the resulting behavior space. In spaces where the majority of controllers are random or uninteresting, it is possible that increasing $k$ will result in more distinct behaviors but would return many medoids that are random, presenting the user with significant multiplicity in the returned behaviors. 
% Future work should consider novel ways to represent the diversity found within the behavior space in a way that minimizes the information returned to the user while still fully representing the behavior space and all behaviors discovered within it. 

\section{Summary}
In this work, we seek to answer the following question: what is the set of possible emergent behaviors in a swarm of robots
possessing a specific set of individual capabilities? To address this question, we present a novel approach that utilizes human feedback to  better understand the collective capability of a computation-free robot swarm in terms of the possible emergent behaviors. We demonstrate the ability for a deep neural network trained with contrastive learning and human similarity queries to replace hand-crafted mappings into low-dimensional behavior representations. This enables the discovery of novel emergent swarm behaviors without requiring any domain expertise in swarm mechanics or explicit behavior metrics. To improve the efficiency of our results, we also contribute a set of general heuristics for filtering out controllers that are unlikely to lead to interesting emergent behaviors.

We present the results of our methods when evolving a set of emergent behaviors and find that we are able to re-discover all known emergent collective behaviors for a computation-free, single-sensor swarm. We also provide the first exploration and analysis of emergent behaviors that are possible give a swarm of computation-free, two-sensor robots. We present two new behaviors, Nested Cycles and Concave Paths, that are possible given a two-sensor robot model. To the best of our knowledge, both behaviors were previously unknown to be possible for computation-free swarms.

%%
%% The acknowledg4.5ments section is defined using the "acks" environment
% \begin{acks}
% \end{acks}

\bibliographystyle{ACM-Reference-Format}
\bibliography{main}


\begin{thebibliography}{49}

%%% the Web address.

\ifx \showCODEN    \undefined \def \showCODEN     #1{\unskip}     \fi
\ifx \showDOI      \undefined \def \showDOI       #1{#1}\fi
\ifx \showISBNx    \undefined \def \showISBNx     #1{\unskip}     \fi
\ifx \showISBNxiii \undefined \def \showISBNxiii  #1{\unskip}     \fi
\ifx \showISSN     \undefined \def \showISSN      #1{\unskip}     \fi
\ifx \showLCCN     \undefined \def \showLCCN      #1{\unskip}     \fi
\ifx \shownote     \undefined \def \shownote      #1{#1}          \fi
\ifx \showarticletitle \undefined \def \showarticletitle #1{#1}   \fi
\ifx \showURL      \undefined \def \showURL       {\relax}        \fi
% invisible to TeX
\providecommand\bibfield[2]{#2}
\providecommand\bibinfo[2]{#2}
\providecommand\natexlab[1]{#1}
\providecommand\showeprint[2][]{arXiv:#2}

\bibitem[\protect\citeauthoryear{??}{pyg}{[n.\,d.]}]%
        {pygame.org}
 \bibinfo{year}{[n.\,d.]}\natexlab{}.
\newblock \bibinfo{title}{Pygame}.
\newblock
\newblock
\urldef\tempurl%
\url{https://www.pygame.org/}
\showURL{%
\tempurl}


\bibitem[\protect\citeauthoryear{Adamatzky}{Adamatzky}{2010}]%
        {adamatzky2010game}
\bibfield{author}{\bibinfo{person}{Andrew Adamatzky}.}
  \bibinfo{year}{2010}\natexlab{}.
\newblock \bibinfo{booktitle}{\emph{Game of life cellular automata}}.
  Vol.~\bibinfo{volume}{1}.
\newblock \bibinfo{publisher}{Springer}.
\newblock


\bibitem[\protect\citeauthoryear{Agarwal, Wills, Cayton, Lanckriet, Kriegman,
  and Belongie}{Agarwal et~al\mbox{.}}{2007}]%
        {agarwal2007generalized}
\bibfield{author}{\bibinfo{person}{Sameer Agarwal}, \bibinfo{person}{Josh
  Wills}, \bibinfo{person}{Lawrence Cayton}, \bibinfo{person}{Gert Lanckriet},
  \bibinfo{person}{David Kriegman}, {and} \bibinfo{person}{Serge Belongie}.}
  \bibinfo{year}{2007}\natexlab{}.
\newblock \showarticletitle{Generalized non-metric multidimensional scaling}.
  In \bibinfo{booktitle}{\emph{Artificial Intelligence and Statistics}}. PMLR,
  \bibinfo{pages}{11--18}.
\newblock


\bibitem[\protect\citeauthoryear{Al-Halah, Ramakrishnan, and Grauman}{Al-Halah
  et~al\mbox{.}}{2022}]%
        {al2022zero}
\bibfield{author}{\bibinfo{person}{Ziad Al-Halah},
  \bibinfo{person}{Santhosh~Kumar Ramakrishnan}, {and} \bibinfo{person}{Kristen
  Grauman}.} \bibinfo{year}{2022}\natexlab{}.
\newblock \showarticletitle{Zero experience required: Plug \& play modular
  transfer learning for semantic visual navigation}. In
  \bibinfo{booktitle}{\emph{Proceedings of the IEEE/CVF Conference on Computer
  Vision and Pattern Recognition}}. \bibinfo{pages}{17031--17041}.
\newblock


\bibitem[\protect\citeauthoryear{Amid, Gionis, and Ukkonen}{Amid
  et~al\mbox{.}}{2015}]%
        {ehsan2015kernel}
\bibfield{author}{\bibinfo{person}{Ehsan Amid}, \bibinfo{person}{Aristides
  Gionis}, {and} \bibinfo{person}{Antti Ukkonen}.}
  \bibinfo{year}{2015}\natexlab{}.
\newblock \showarticletitle{A Kernel-Learning Approach to Semi-supervised
  Clustering with Relative Distance Comparisons}, Vol.~\bibinfo{volume}{9284}.
\newblock
\showISBNx{978-3-319-23527-1}
\urldef\tempurl%
\url{https://doi.org/10.1007/978-3-319-23528-8_14}
\showDOI{\tempurl}


\bibitem[\protect\citeauthoryear{Balntas, Riba, Ponsa, and Mikolajczyk}{Balntas
  et~al\mbox{.}}{2016}]%
        {balntas2016learning}
\bibfield{author}{\bibinfo{person}{Vassileios Balntas}, \bibinfo{person}{Edgar
  Riba}, \bibinfo{person}{Daniel Ponsa}, {and} \bibinfo{person}{Krystian
  Mikolajczyk}.} \bibinfo{year}{2016}\natexlab{}.
\newblock \showarticletitle{Learning local feature descriptors with triplets
  and shallow convolutional neural networks.}. In
  \bibinfo{booktitle}{\emph{Bmvc}}, Vol.~\bibinfo{volume}{1}.
  \bibinfo{pages}{3}.
\newblock


\bibitem[\protect\citeauthoryear{Bayindir and {\c{S}}ahin}{Bayindir and
  {\c{S}}ahin}{2007}]%
        {bayindir2007review}
\bibfield{author}{\bibinfo{person}{Levent Bayindir} {and} \bibinfo{person}{Erol
  {\c{S}}ahin}.} \bibinfo{year}{2007}\natexlab{}.
\newblock \showarticletitle{A review of studies in swarm robotics}.
\newblock \bibinfo{journal}{\emph{Turkish Journal of Electrical Engineering and
  Computer Sciences}} \bibinfo{volume}{15}, \bibinfo{number}{2}
  (\bibinfo{year}{2007}), \bibinfo{pages}{115--147}.
\newblock


\bibitem[\protect\citeauthoryear{Brown, Turner, Hennigh, and Loscalzo}{Brown
  et~al\mbox{.}}{2018}]%
        {brown2018discovery}
\bibfield{author}{\bibinfo{person}{Daniel~S Brown}, \bibinfo{person}{Ryan
  Turner}, \bibinfo{person}{Oliver Hennigh}, {and} \bibinfo{person}{Steven
  Loscalzo}.} \bibinfo{year}{2018}\natexlab{}.
\newblock \showarticletitle{Discovery and exploration of novel swarm behaviors
  given limited robot capabilities}.
\newblock In \bibinfo{booktitle}{\emph{Distributed Autonomous Robotic
  Systems}}. \bibinfo{publisher}{Springer}, \bibinfo{pages}{447--460}.
\newblock


\bibitem[\protect\citeauthoryear{Chen, Kornblith, Norouzi, and Hinton}{Chen
  et~al\mbox{.}}{2020}]%
        {chen2020simple}
\bibfield{author}{\bibinfo{person}{Ting Chen}, \bibinfo{person}{Simon
  Kornblith}, \bibinfo{person}{Mohammad Norouzi}, {and}
  \bibinfo{person}{Geoffrey Hinton}.} \bibinfo{year}{2020}\natexlab{}.
\newblock \showarticletitle{A simple framework for contrastive learning of
  visual representations}. In \bibinfo{booktitle}{\emph{International
  conference on machine learning}}. PMLR, \bibinfo{pages}{1597--1607}.
\newblock


\bibitem[\protect\citeauthoryear{Cliff, Husbands, and Harvey}{Cliff
  et~al\mbox{.}}{1993}]%
        {cliff1993explorations}
\bibfield{author}{\bibinfo{person}{Dave Cliff}, \bibinfo{person}{Phil
  Husbands}, {and} \bibinfo{person}{Inman Harvey}.}
  \bibinfo{year}{1993}\natexlab{}.
\newblock \showarticletitle{Explorations in evolutionary robotics}.
\newblock \bibinfo{journal}{\emph{Adaptive behavior}} \bibinfo{volume}{2},
  \bibinfo{number}{1} (\bibinfo{year}{1993}), \bibinfo{pages}{73--110}.
\newblock


\bibitem[\protect\citeauthoryear{Demiralp, Bernstein, and Heer}{Demiralp
  et~al\mbox{.}}{2014}]%
        {cagatay2014kernels}
\bibfield{author}{\bibinfo{person}{Cagatay Demiralp}, \bibinfo{person}{Michael
  Bernstein}, {and} \bibinfo{person}{Jeffrey Heer}.}
  \bibinfo{year}{2014}\natexlab{}.
\newblock \showarticletitle{Learning Perceptual Kernels for Visualization
  Design}.
\newblock \bibinfo{journal}{\emph{IEEE Transactions on Visualization and
  Computer Graphics}}  \bibinfo{volume}{20}.
\newblock
\urldef\tempurl%
\url{https://doi.org/10.1109/TVCG.2014.2346978}
\showDOI{\tempurl}


\bibitem[\protect\citeauthoryear{Dias, Silva, Rocha~Filho, Vargas, Cota, and
  Pessin}{Dias et~al\mbox{.}}{2021}]%
        {dias2021swarm}
\bibfield{author}{\bibinfo{person}{Pollyanna G~Faria Dias},
  \bibinfo{person}{Mateus~C Silva}, \bibinfo{person}{Geraldo~P Rocha~Filho},
  \bibinfo{person}{Patr{\'\i}cia~A Vargas}, \bibinfo{person}{Luciano~P Cota},
  {and} \bibinfo{person}{Gustavo Pessin}.} \bibinfo{year}{2021}\natexlab{}.
\newblock \showarticletitle{Swarm robotics: A perspective on the latest
  reviewed concepts and applications}.
\newblock \bibinfo{journal}{\emph{Sensors}} \bibinfo{volume}{21},
  \bibinfo{number}{6} (\bibinfo{year}{2021}), \bibinfo{pages}{2062}.
\newblock


\bibitem[\protect\citeauthoryear{Dorigo, Theraulaz, and Trianni}{Dorigo
  et~al\mbox{.}}{2021}]%
        {dorigo2021swarm}
\bibfield{author}{\bibinfo{person}{Marco Dorigo}, \bibinfo{person}{Guy
  Theraulaz}, {and} \bibinfo{person}{Vito Trianni}.}
  \bibinfo{year}{2021}\natexlab{}.
\newblock \showarticletitle{Swarm robotics: Past, present, and future [point of
  view]}.
\newblock \bibinfo{journal}{\emph{Proc. IEEE}} \bibinfo{volume}{109},
  \bibinfo{number}{7} (\bibinfo{year}{2021}), \bibinfo{pages}{1152--1165}.
\newblock


\bibitem[\protect\citeauthoryear{Engebr{\aa}ten, Moen, Yakimenko, and
  Glette}{Engebr{\aa}ten et~al\mbox{.}}{2018}]%
        {engebraaten2018evolving}
\bibfield{author}{\bibinfo{person}{Sondre~A Engebr{\aa}ten},
  \bibinfo{person}{Jonas Moen}, \bibinfo{person}{Oleg Yakimenko}, {and}
  \bibinfo{person}{Kyrre Glette}.} \bibinfo{year}{2018}\natexlab{}.
\newblock \showarticletitle{Evolving a repertoire of controllers for a
  multi-function swarm}. In \bibinfo{booktitle}{\emph{Applications of
  Evolutionary Computation: 21st International Conference, EvoApplications
  2018, Parma, Italy, April 4-6, 2018, Proceedings 21}}. Springer,
  \bibinfo{pages}{734--749}.
\newblock


\bibitem[\protect\citeauthoryear{Fontaine and Nikolaidis}{Fontaine and
  Nikolaidis}{2021}]%
        {fontaine2021differentiable}
\bibfield{author}{\bibinfo{person}{Matthew Fontaine} {and}
  \bibinfo{person}{Stefanos Nikolaidis}.} \bibinfo{year}{2021}\natexlab{}.
\newblock \showarticletitle{Differentiable quality diversity}.
\newblock \bibinfo{journal}{\emph{Advances in Neural Information Processing
  Systems}}  \bibinfo{volume}{34} (\bibinfo{year}{2021}),
  \bibinfo{pages}{10040--10052}.
\newblock


\bibitem[\protect\citeauthoryear{Gauci, Chen, Dodd, and Gro{\ss}}{Gauci
  et~al\mbox{.}}{2014a}]%
        {gauci2014evolving}
\bibfield{author}{\bibinfo{person}{Melvin Gauci}, \bibinfo{person}{Jianing
  Chen}, \bibinfo{person}{Tony~J Dodd}, {and} \bibinfo{person}{Roderich
  Gro{\ss}}.} \bibinfo{year}{2014}\natexlab{a}.
\newblock \showarticletitle{Evolving aggregation behaviors in multi-robot
  systems with binary sensors}.
\newblock In \bibinfo{booktitle}{\emph{Distributed autonomous robotic
  systems}}. \bibinfo{publisher}{Springer}, \bibinfo{pages}{355--367}.
\newblock


\bibitem[\protect\citeauthoryear{Gauci, Chen, Li, Dodd, and Gro{\ss}}{Gauci
  et~al\mbox{.}}{2014b}]%
        {gauci2014clustering}
\bibfield{author}{\bibinfo{person}{Melvin Gauci}, \bibinfo{person}{Jianing
  Chen}, \bibinfo{person}{Wei Li}, \bibinfo{person}{Tony~J Dodd}, {and}
  \bibinfo{person}{Roderich Gro{\ss}}.} \bibinfo{year}{2014}\natexlab{b}.
\newblock \showarticletitle{Clustering objects with robots that do not
  compute}. In \bibinfo{booktitle}{\emph{Proceedings of the 2014 international
  conference on Autonomous agents and multi-agent systems}}.
  \bibinfo{pages}{421--428}.
\newblock


\bibitem[\protect\citeauthoryear{Gomes and Christensen}{Gomes and
  Christensen}{2013}]%
        {gomes2013generic}
\bibfield{author}{\bibinfo{person}{Jorge Gomes} {and} \bibinfo{person}{Anders~L
  Christensen}.} \bibinfo{year}{2013}\natexlab{}.
\newblock \showarticletitle{Generic behaviour similarity measures for
  evolutionary swarm robotics}. In \bibinfo{booktitle}{\emph{Proceedings of the
  15th annual conference on Genetic and evolutionary computation}}.
  \bibinfo{pages}{199--206}.
\newblock


\bibitem[\protect\citeauthoryear{Gomes, Mariano, and Christensen}{Gomes
  et~al\mbox{.}}{2014}]%
        {gomes2014systematic}
\bibfield{author}{\bibinfo{person}{Jorge Gomes}, \bibinfo{person}{Pedro
  Mariano}, {and} \bibinfo{person}{Anders Christensen}.}
  \bibinfo{year}{2014}\natexlab{}.
\newblock \showarticletitle{Systematic Derivation of Behaviour
  Characterisations in Evolutionary Robotics}. In
  \bibinfo{booktitle}{\emph{Artificial Life Conference Proceedings}}. MIT Press
  One Rogers Street, Cambridge, MA 02142-1209, USA journals-info~…,
  \bibinfo{pages}{212--219}.
\newblock


\bibitem[\protect\citeauthoryear{Gomes, Oliveira, and Christensen}{Gomes
  et~al\mbox{.}}{2018}]%
        {gomes2018approach}
\bibfield{author}{\bibinfo{person}{Jorge Gomes}, \bibinfo{person}{Sancho~Moura
  Oliveira}, {and} \bibinfo{person}{Anders~Lyhne Christensen}.}
  \bibinfo{year}{2018}\natexlab{}.
\newblock \showarticletitle{An approach to evolve and exploit repertoires of
  general robot behaviours}.
\newblock \bibinfo{journal}{\emph{Swarm and Evolutionary Computation}}
  \bibinfo{volume}{43} (\bibinfo{year}{2018}), \bibinfo{pages}{265--283}.
\newblock


\bibitem[\protect\citeauthoryear{Gravina, Liapis, and Yannakakis}{Gravina
  et~al\mbox{.}}{2018}]%
        {gravina2018quality}
\bibfield{author}{\bibinfo{person}{Daniele Gravina}, \bibinfo{person}{Antonios
  Liapis}, {and} \bibinfo{person}{Georgios~N Yannakakis}.}
  \bibinfo{year}{2018}\natexlab{}.
\newblock \showarticletitle{Quality diversity through surprise}.
\newblock \bibinfo{journal}{\emph{IEEE Transactions on Evolutionary
  Computation}} \bibinfo{volume}{23}, \bibinfo{number}{4}
  (\bibinfo{year}{2018}), \bibinfo{pages}{603--616}.
\newblock


\bibitem[\protect\citeauthoryear{Grillotti and Cully}{Grillotti and
  Cully}{2022}]%
        {grillotti2022unsupervised}
\bibfield{author}{\bibinfo{person}{Luca Grillotti} {and}
  \bibinfo{person}{Antoine Cully}.} \bibinfo{year}{2022}\natexlab{}.
\newblock \showarticletitle{Unsupervised Behavior Discovery With
  Quality-Diversity Optimization}.
\newblock \bibinfo{journal}{\emph{IEEE Transactions on Evolutionary
  Computation}} \bibinfo{volume}{26}, \bibinfo{number}{6}
  (\bibinfo{year}{2022}), \bibinfo{pages}{1539--1552}.
\newblock


\bibitem[\protect\citeauthoryear{Hadsell, Chopra, and LeCun}{Hadsell
  et~al\mbox{.}}{2006}]%
        {hadsell2006dimensionality}
\bibfield{author}{\bibinfo{person}{Raia Hadsell}, \bibinfo{person}{Sumit
  Chopra}, {and} \bibinfo{person}{Yann LeCun}.}
  \bibinfo{year}{2006}\natexlab{}.
\newblock \showarticletitle{Dimensionality reduction by learning an invariant
  mapping}. In \bibinfo{booktitle}{\emph{2006 IEEE Computer Society Conference
  on Computer Vision and Pattern Recognition (CVPR'06)}},
  Vol.~\bibinfo{volume}{2}. IEEE, \bibinfo{pages}{1735--1742}.
\newblock


\bibitem[\protect\citeauthoryear{Hoffer and Ailon}{Hoffer and Ailon}{2015}]%
        {hoffer2015deep}
\bibfield{author}{\bibinfo{person}{Elad Hoffer} {and} \bibinfo{person}{Nir
  Ailon}.} \bibinfo{year}{2015}\natexlab{}.
\newblock \showarticletitle{Deep metric learning using triplet network}. In
  \bibinfo{booktitle}{\emph{International workshop on similarity-based pattern
  recognition}}. Springer, \bibinfo{pages}{84--92}.
\newblock


\bibitem[\protect\citeauthoryear{Hoffman, Gelman, et~al\mbox{.}}{Hoffman
  et~al\mbox{.}}{2014}]%
        {hoffman2014no}
\bibfield{author}{\bibinfo{person}{Matthew~D Hoffman}, \bibinfo{person}{Andrew
  Gelman}, {et~al\mbox{.}}} \bibinfo{year}{2014}\natexlab{}.
\newblock \showarticletitle{The No-U-Turn sampler: adaptively setting path
  lengths in Hamiltonian Monte Carlo.}
\newblock \bibinfo{journal}{\emph{J. Mach. Learn. Res.}} \bibinfo{volume}{15},
  \bibinfo{number}{1} (\bibinfo{year}{2014}), \bibinfo{pages}{1593--1623}.
\newblock


\bibitem[\protect\citeauthoryear{Jaiswal, Babu, Zadeh, Banerjee, and
  Makedon}{Jaiswal et~al\mbox{.}}{2020}]%
        {jaiswal2020survey}
\bibfield{author}{\bibinfo{person}{Ashish Jaiswal},
  \bibinfo{person}{Ashwin~Ramesh Babu}, \bibinfo{person}{Mohammad~Zaki Zadeh},
  \bibinfo{person}{Debapriya Banerjee}, {and} \bibinfo{person}{Fillia
  Makedon}.} \bibinfo{year}{2020}\natexlab{}.
\newblock \showarticletitle{A survey on contrastive self-supervised learning}.
\newblock \bibinfo{journal}{\emph{Technologies}} \bibinfo{volume}{9},
  \bibinfo{number}{1} (\bibinfo{year}{2020}), \bibinfo{pages}{2}.
\newblock


\bibitem[\protect\citeauthoryear{Johnson and Khoshgoftaar}{Johnson and
  Khoshgoftaar}{2019}]%
        {johnson2019survey}
\bibfield{author}{\bibinfo{person}{Justin~M Johnson} {and}
  \bibinfo{person}{Taghi~M Khoshgoftaar}.} \bibinfo{year}{2019}\natexlab{}.
\newblock \showarticletitle{Survey on deep learning with class imbalance}.
\newblock \bibinfo{journal}{\emph{Journal of Big Data}} \bibinfo{volume}{6},
  \bibinfo{number}{1} (\bibinfo{year}{2019}), \bibinfo{pages}{1--54}.
\newblock


\bibitem[\protect\citeauthoryear{Johnson and Brown}{Johnson and Brown}{2016}]%
        {johnson2016evolving}
\bibfield{author}{\bibinfo{person}{Matthew Johnson} {and}
  \bibinfo{person}{Daniel Brown}.} \bibinfo{year}{2016}\natexlab{}.
\newblock \showarticletitle{Evolving and controlling perimeter, rendezvous, and
  foraging behaviors in a computation-free robot swarm}. In
  \bibinfo{booktitle}{\emph{Proceedings of the 9th EAI International Conference
  on Bio-inspired Information and Communications Technologies (formerly
  BIONETICS)}}. \bibinfo{pages}{311--314}.
\newblock


\bibitem[\protect\citeauthoryear{Lehman and Stanley}{Lehman and
  Stanley}{2011}]%
        {lehman2011abandoning}
\bibfield{author}{\bibinfo{person}{Joel Lehman} {and}
  \bibinfo{person}{Kenneth~O Stanley}.} \bibinfo{year}{2011}\natexlab{}.
\newblock \showarticletitle{Abandoning objectives: Evolution through the search
  for novelty alone}.
\newblock \bibinfo{journal}{\emph{Evolutionary computation}}
  \bibinfo{volume}{19}, \bibinfo{number}{2} (\bibinfo{year}{2011}),
  \bibinfo{pages}{189--223}.
\newblock


\bibitem[\protect\citeauthoryear{Liapis, Mart{\'\i}nez, Togelius, and
  Yannakakis}{Liapis et~al\mbox{.}}{2021}]%
        {liapis2021transforming}
\bibfield{author}{\bibinfo{person}{Antonios Liapis},
  \bibinfo{person}{H{\'e}ctor~P Mart{\'\i}nez}, \bibinfo{person}{Julian
  Togelius}, {and} \bibinfo{person}{Georgios~N Yannakakis}.}
  \bibinfo{year}{2021}\natexlab{}.
\newblock \showarticletitle{Transforming exploratory creativity with DeLeNoX}.
\newblock \bibinfo{journal}{\emph{arXiv preprint arXiv:2103.11715}}
  (\bibinfo{year}{2021}).
\newblock


\bibitem[\protect\citeauthoryear{McFee, Lanckriet, and Jebara}{McFee
  et~al\mbox{.}}{2011}]%
        {mcfee2011learning}
\bibfield{author}{\bibinfo{person}{Brian McFee}, \bibinfo{person}{Gert
  Lanckriet}, {and} \bibinfo{person}{Tony Jebara}.}
  \bibinfo{year}{2011}\natexlab{}.
\newblock \showarticletitle{Learning Multi-modal Similarity.}
\newblock \bibinfo{journal}{\emph{Journal of machine learning research}}
  \bibinfo{volume}{12}, \bibinfo{number}{2} (\bibinfo{year}{2011}).
\newblock


\bibitem[\protect\citeauthoryear{Meyerson, Lehman, and Miikkulainen}{Meyerson
  et~al\mbox{.}}{2016}]%
        {meyerson2016learning}
\bibfield{author}{\bibinfo{person}{Elliot Meyerson}, \bibinfo{person}{Joel
  Lehman}, {and} \bibinfo{person}{Risto Miikkulainen}.}
  \bibinfo{year}{2016}\natexlab{}.
\newblock \showarticletitle{Learning behavior characterizations for novelty
  search}. In \bibinfo{booktitle}{\emph{Proceedings of the Genetic and
  Evolutionary Computation Conference 2016}}. \bibinfo{pages}{149--156}.
\newblock


\bibitem[\protect\citeauthoryear{Mirhosseini, Zion, Dauchot, and
  Bredeche}{Mirhosseini et~al\mbox{.}}{2022}]%
        {MirhosseiniAdaptive}
\bibfield{author}{\bibinfo{person}{Yoones Mirhosseini}, \bibinfo{person}{Matan
  Yah~Ben Zion}, \bibinfo{person}{Olivier Dauchot}, {and}
  \bibinfo{person}{Nicolas Bredeche}.} \bibinfo{year}{2022}\natexlab{}.
\newblock \showarticletitle{Adaptive Phototaxis of a Swarm of Mobile Robots
  Using Positive and Negative Feedback Self-Alignment}. In
  \bibinfo{booktitle}{\emph{Proceedings of the Genetic and Evolutionary
  Computation Conference}} (Boston, Massachusetts)
  \emph{(\bibinfo{series}{GECCO '22})}. \bibinfo{publisher}{Association for
  Computing Machinery}, \bibinfo{address}{New York, NY, USA},
  \bibinfo{pages}{104–112}.
\newblock
\showISBNx{9781450392372}
\urldef\tempurl%
\url{https://doi.org/10.1145/3512290.3528816}
\showDOI{\tempurl}


\bibitem[\protect\citeauthoryear{Mouret and Doncieux}{Mouret and
  Doncieux}{2012}]%
        {mouret2012encouraging}
\bibfield{author}{\bibinfo{person}{J-B Mouret} {and}
  \bibinfo{person}{St{\'e}phane Doncieux}.} \bibinfo{year}{2012}\natexlab{}.
\newblock \showarticletitle{Encouraging behavioral diversity in evolutionary
  robotics: An empirical study}.
\newblock \bibinfo{journal}{\emph{Evolutionary computation}}
  \bibinfo{volume}{20}, \bibinfo{number}{1} (\bibinfo{year}{2012}),
  \bibinfo{pages}{91--133}.
\newblock


\bibitem[\protect\citeauthoryear{Neal et~al\mbox{.}}{Neal
  et~al\mbox{.}}{2011}]%
        {neal2011mcmc}
\bibfield{author}{\bibinfo{person}{Radford~M Neal} {et~al\mbox{.}}}
  \bibinfo{year}{2011}\natexlab{}.
\newblock \showarticletitle{MCMC using Hamiltonian dynamics}.
\newblock \bibinfo{journal}{\emph{Handbook of markov chain monte carlo}}
  \bibinfo{volume}{2}, \bibinfo{number}{11} (\bibinfo{year}{2011}),
  \bibinfo{pages}{2}.
\newblock


\bibitem[\protect\citeauthoryear{Nguyen, Yosinski, and Clune}{Nguyen
  et~al\mbox{.}}{2015}]%
        {nguyen2015innovation}
\bibfield{author}{\bibinfo{person}{Anh~Mai Nguyen}, \bibinfo{person}{Jason
  Yosinski}, {and} \bibinfo{person}{Jeff Clune}.}
  \bibinfo{year}{2015}\natexlab{}.
\newblock \showarticletitle{Innovation engines: Automated creativity and
  improved stochastic optimization via deep learning}. In
  \bibinfo{booktitle}{\emph{Proceedings of the 2015 annual conference on
  genetic and evolutionary computation}}. \bibinfo{pages}{959--966}.
\newblock


\bibitem[\protect\citeauthoryear{Nolfi and Floreano}{Nolfi and
  Floreano}{2000}]%
        {nolfi2000evolutionary}
\bibfield{author}{\bibinfo{person}{Stefano Nolfi} {and} \bibinfo{person}{Dario
  Floreano}.} \bibinfo{year}{2000}\natexlab{}.
\newblock \bibinfo{booktitle}{\emph{Evolutionary robotics: The biology,
  intelligence, and technology of self-organizing machines}}.
\newblock \bibinfo{publisher}{MIT press}.
\newblock


\bibitem[\protect\citeauthoryear{{\"O}zdemir, Gauci, and Gro{\ss}}{{\"O}zdemir
  et~al\mbox{.}}{2017}]%
        {ozdemir2017shepherding}
\bibfield{author}{\bibinfo{person}{Anil {\"O}zdemir}, \bibinfo{person}{Melvin
  Gauci}, {and} \bibinfo{person}{Roderich Gro{\ss}}.}
  \bibinfo{year}{2017}\natexlab{}.
\newblock \showarticletitle{Shepherding with robots that do not compute}. In
  \bibinfo{booktitle}{\emph{ECAL 2017, the Fourteenth European Conference on
  Artificial Life}}. MIT Press, \bibinfo{pages}{332--339}.
\newblock


\bibitem[\protect\citeauthoryear{{\"O}zdemir, Gauci, Kolling, Hall, and
  Gro{\ss}}{{\"O}zdemir et~al\mbox{.}}{2019}]%
        {ozdemir2019spatial}
\bibfield{author}{\bibinfo{person}{An{\i}l {\"O}zdemir},
  \bibinfo{person}{Melvin Gauci}, \bibinfo{person}{Andreas Kolling},
  \bibinfo{person}{Matthew~D Hall}, {and} \bibinfo{person}{Roderich Gro{\ss}}.}
  \bibinfo{year}{2019}\natexlab{}.
\newblock \showarticletitle{Spatial coverage without computation}. In
  \bibinfo{booktitle}{\emph{2019 International Conference on Robotics and
  Automation (ICRA)}}. IEEE, \bibinfo{pages}{9674--9680}.
\newblock


\bibitem[\protect\citeauthoryear{Parrish and Edelstein-Keshet}{Parrish and
  Edelstein-Keshet}{1999}]%
        {parrish1999complexity}
\bibfield{author}{\bibinfo{person}{Julia~K Parrish} {and} \bibinfo{person}{Leah
  Edelstein-Keshet}.} \bibinfo{year}{1999}\natexlab{}.
\newblock \showarticletitle{Complexity, pattern, and evolutionary trade-offs in
  animal aggregation}.
\newblock \bibinfo{journal}{\emph{Science}} \bibinfo{volume}{284},
  \bibinfo{number}{5411} (\bibinfo{year}{1999}), \bibinfo{pages}{99--101}.
\newblock


\bibitem[\protect\citeauthoryear{Pugh, Soros, and Stanley}{Pugh
  et~al\mbox{.}}{2016}]%
        {pugh2016quality}
\bibfield{author}{\bibinfo{person}{Justin~K Pugh}, \bibinfo{person}{Lisa~B
  Soros}, {and} \bibinfo{person}{Kenneth~O Stanley}.}
  \bibinfo{year}{2016}\natexlab{}.
\newblock \showarticletitle{Quality diversity: A new frontier for evolutionary
  computation}.
\newblock \bibinfo{journal}{\emph{Frontiers in Robotics and AI}}
  (\bibinfo{year}{2016}), \bibinfo{pages}{40}.
\newblock


\bibitem[\protect\citeauthoryear{Sayama}{Sayama}{2009}]%
        {sayama2009swarm}
\bibfield{author}{\bibinfo{person}{Hiroki Sayama}.}
  \bibinfo{year}{2009}\natexlab{}.
\newblock \showarticletitle{Swarm chemistry}.
\newblock \bibinfo{journal}{\emph{Artificial life}} \bibinfo{volume}{15},
  \bibinfo{number}{1} (\bibinfo{year}{2009}), \bibinfo{pages}{105--114}.
\newblock


\bibitem[\protect\citeauthoryear{Sperati, Trianni, and Nolfi}{Sperati
  et~al\mbox{.}}{2011}]%
        {sperati2011self}
\bibfield{author}{\bibinfo{person}{Valerio Sperati}, \bibinfo{person}{Vito
  Trianni}, {and} \bibinfo{person}{Stefano Nolfi}.}
  \bibinfo{year}{2011}\natexlab{}.
\newblock \showarticletitle{Self-organised path formation in a swarm of
  robots}.
\newblock \bibinfo{journal}{\emph{Swarm Intelligence}}  \bibinfo{volume}{5}
  (\bibinfo{year}{2011}), \bibinfo{pages}{97--119}.
\newblock


\bibitem[\protect\citeauthoryear{St-Onge, Pinciroli, and Beltrame}{St-Onge
  et~al\mbox{.}}{2018}]%
        {st2018circle}
\bibfield{author}{\bibinfo{person}{David St-Onge}, \bibinfo{person}{Carlo
  Pinciroli}, {and} \bibinfo{person}{Giovanni Beltrame}.}
  \bibinfo{year}{2018}\natexlab{}.
\newblock \showarticletitle{Circle formation with computation-free robots shows
  emergent behavioural structure}. In \bibinfo{booktitle}{\emph{2018 IEEE/RSJ
  International Conference on Intelligent Robots and Systems (IROS)}}. IEEE,
  \bibinfo{pages}{5344--5349}.
\newblock


\bibitem[\protect\citeauthoryear{Stolfi and Danoy}{Stolfi and Danoy}{2022}]%
        {StolfiAutonomous}
\bibfield{author}{\bibinfo{person}{Daniel~H. Stolfi} {and}
  \bibinfo{person}{Gr\'{e}goire Danoy}.} \bibinfo{year}{2022}\natexlab{}.
\newblock \showarticletitle{Optimising Autonomous Robot Swarm Parameters for
  Stable Formation Design}. In \bibinfo{booktitle}{\emph{Proceedings of the
  Genetic and Evolutionary Computation Conference}} (Boston, Massachusetts)
  \emph{(\bibinfo{series}{GECCO '22})}. \bibinfo{publisher}{Association for
  Computing Machinery}, \bibinfo{address}{New York, NY, USA},
  \bibinfo{pages}{1281–1289}.
\newblock
\showISBNx{9781450392372}
\urldef\tempurl%
\url{https://doi.org/10.1145/3512290.3528709}
\showDOI{\tempurl}


\bibitem[\protect\citeauthoryear{Tamuz, Liu, Belongie, Shamir, and Kalai}{Tamuz
  et~al\mbox{.}}{2011}]%
        {tamuz2011adaptively}
\bibfield{author}{\bibinfo{person}{Omer Tamuz}, \bibinfo{person}{Ce Liu},
  \bibinfo{person}{Serge Belongie}, \bibinfo{person}{Ohad Shamir}, {and}
  \bibinfo{person}{Adam~Tauman Kalai}.} \bibinfo{year}{2011}\natexlab{}.
\newblock \showarticletitle{Adaptively learning the crowd kernel}.
\newblock \bibinfo{journal}{\emph{arXiv preprint arXiv:1105.1033}}
  (\bibinfo{year}{2011}).
\newblock


\bibitem[\protect\citeauthoryear{Trianni, Gro{\ss}, Labella, {\c{S}}ahin, and
  Dorigo}{Trianni et~al\mbox{.}}{2003}]%
        {trianni2003evolving}
\bibfield{author}{\bibinfo{person}{Vito Trianni}, \bibinfo{person}{Roderich
  Gro{\ss}}, \bibinfo{person}{Thomas~H Labella}, \bibinfo{person}{Erol
  {\c{S}}ahin}, {and} \bibinfo{person}{Marco Dorigo}.}
  \bibinfo{year}{2003}\natexlab{}.
\newblock \showarticletitle{Evolving aggregation behaviors in a swarm of
  robots}. In \bibinfo{booktitle}{\emph{Advances in Artificial Life: 7th
  European Conference, ECAL 2003, Dortmund, Germany, September 14-17, 2003.
  Proceedings 7}}. Springer, \bibinfo{pages}{865--874}.
\newblock


\bibitem[\protect\citeauthoryear{Van~der Maaten and Hinton}{Van~der Maaten and
  Hinton}{2008}]%
        {van2008visualizing}
\bibfield{author}{\bibinfo{person}{Laurens Van~der Maaten} {and}
  \bibinfo{person}{Geoffrey Hinton}.} \bibinfo{year}{2008}\natexlab{}.
\newblock \showarticletitle{Visualizing data using t-SNE.}
\newblock \bibinfo{journal}{\emph{Journal of machine learning research}}
  \bibinfo{volume}{9}, \bibinfo{number}{11} (\bibinfo{year}{2008}).
\newblock


\bibitem[\protect\citeauthoryear{Zhai, Oliver, Kolesnikov, and Beyer}{Zhai
  et~al\mbox{.}}{2019}]%
        {zhai2019s4l}
\bibfield{author}{\bibinfo{person}{Xiaohua Zhai}, \bibinfo{person}{Avital
  Oliver}, \bibinfo{person}{Alexander Kolesnikov}, {and} \bibinfo{person}{Lucas
  Beyer}.} \bibinfo{year}{2019}\natexlab{}.
\newblock \showarticletitle{S4l: Self-supervised semi-supervised learning}. In
  \bibinfo{booktitle}{\emph{Proceedings of the IEEE/CVF International
  Conference on Computer Vision}}. \bibinfo{pages}{1476--1485}.
\newblock


\end{thebibliography}

\newpage
\appendix

\section{Simulation}
Our custom Swarm Simulator provides a friction-less environment in which we simulate robot interaction. Our simulator makes the assumption that sensing and actuation are perfect and that collisions between agents result in no damage to the agent's structure or capabilities.

\subsection{Swarm Kinematics} \label{Appendix: Swarm Kinematics}
For a single Differential Drive agent with left and right wheel velocities $v_l$ and $v_r$, respectively, and heading $\theta$, we compute the change in position as

\begin{equation}
    \Delta x = \frac{\text{W}_{rad}}{2}(v_l + v_r)cos(\theta)\Delta t,
    \label{eq: kinematics x}
\end{equation}

\begin{equation}
    \Delta y = \frac{\text{W}_{rad}}{2}(v_l + v_r)sin(\theta)\Delta t,
    \label{eq: kinematics y}
\end{equation}

\begin{equation}
    \Delta \theta = \frac{(v_l - v_r)}{2 A_{rad}}\Delta t,
    \label{eq: kinematics theta}
\end{equation}
where $W_{rad}$ is the radius of the wheels, and $A_{rad}$ is the radius of our agents (distance from center to wheel). Each equation operates with respect to a time-step, $\Delta t$. Our experiments use $W_{rad} = 2.0$, $A_{rad} = 5.0$ and $\Delta t = 1.0$ for agents that are 10 units in length. We assume that our binary sensors have infinite range and perfect accuracy.

For a given single-sensor agent at time $t$, we conditionally apply velocities to the left and right wheels based on whether an agents senses another agent or not, $S(t)$. Given the controller from Eq.~\ref{eq: baseline controller}, $C_{1} = [v_1, v_2, v_3, v_4],$ let

\begin{equation}
    v_l(t), v_r(t) = \left\{
        \begin{array}{lr}
            v_1, v_2 & \text{if } S(t) \text{ is "off" }\\
            v_3, v_4 & \text{if } S(t) \text{ is "on" }
        \end{array}
    \right\}. 
    \label{eq:conditional-single-sensor-velocities}
\end{equation}

As a result of adding an additional binary sensor in the two-sensor capability model, the number of conditional branches doubles. For sensors $S_1, S_2$, we determine wheel velocities for the controller $C_{2} = [v_1, v_2, v_3, v_4, v_5, v_6, v_7, v_8, \theta]$ by

\begin{equation}
    v_l(t), v_r(t) = \left\{
        \begin{array}{lr}
            v_1, v_2 & \text{if } S_1(t) \text{ and } S_2(t) \text{ are "off" }\\
            v_3, v_4 & \text{if } S_1(t) \text{ is "on" and } S_2(t) \text{ is "off" }\\
            v_5, v_6 & \text{if } S_1(t) \text{ is "off" and } S_2(t) \text{ is "on" }\\
            v_7, v_8 & \text{if } S_1(t) \text{ and } S_2(t) \text{ are "on" }\\
        \end{array}
    \right\}. 
    \label{eq:conditional-two-sensor-velocities}
\end{equation}

For every time-step in simulation, our simulator calculates the appropriate wheel velocities for every member of the swarm based on sensor data and then executes state updates by calculating equations \ref{eq: kinematics x}-\ref{eq: kinematics theta}.

% \subsection{Horizon Selection and Number of Agents}

\section{Informed Heuristic Filtering} \label{Appendix: Informed Heuristics}
\begin{figure}[h]
     \centering
     \includegraphics[width=\linewidth]{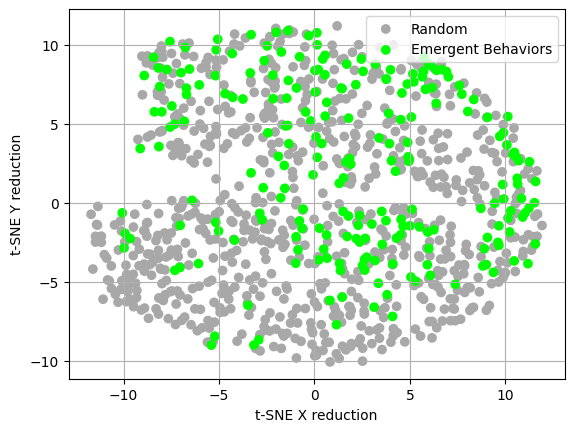}
     \caption{A t-SNE reduction of the single-sensor controller space for a labeled dataset of 1000 behavior classes. Areas of the plane that contain no emergent behaviors serve as a justification that filtering the controller space would reduce the number the search space without compromising behaviors.}
    \label{fig:random controller regions}
\end{figure}

The controller space and it's resulting behavior space contain many random, uninteresting, behaviors. Fig.~\ref{fig:random controller regions} shows a t-SNE reduction of the 4 dimensional controller space to the 2D plane. Although it is difficult to distinguish between random and interesting controllers that are in the same area in this space, there are several regions that are predominantly populated by random behaviors. If we could develop an understanding of why there are some areas of the controller space that do not contain interesting behavior, we could prevent sampling data from there.

In addition to our heuristic, we discretize the $[-1, 1]^4$ controller space into $\{-1.0$,$ -0.9$, $...$, $0.9$, $1.0\}^4$, capping the size of our controller space at $21^4 = 194,481$. 
Similarly, on our two sensor model, the branching factor for decision making (by adding a single sensor) increases the required number of values to 8, with an additional value to represent the sensor angle $\theta \in [-\frac{2\pi}{3}, \frac{2\pi}{3}]$, which we discretize to 10 values $\{\pm\frac{2\pi}{3}$, $\pm\frac{\pi}{2}$, $\pm\frac{\pi}{3}$, $\pm\frac{\pi}{4}$, $\pm\frac{\pi}{6}\}$. By increasing our computation requirements by 1 bit and allowing evolution to play a factor into the design of our agents by adding the placement of the second sensor to our controller, we increase the size of our controller space to $21^8 \times 10 \approx 3.77x10^{12}$ controllers.

Consider the controller spaces for our models represented by Equations \ref{eq: baseline controller} and \ref{eq: two sensor controller}. Brown et al. \cite{brown2018discovery} develop intuition about the behavior of a swarm based on the physics of simulation. For example, the Average Speed of agents is calculated at runtime for a mapping into Brown et al.'s behavior space. This method requires expensive simulation time and fails to utilize information that is known about the controller of the swarm before executing any commands.

\begin{displaymath}
    C_{A} = [0.1, 0.12, 0.25, 0.2], \text{  }  C_{B} = [0.9, 0.98, 0.7, 0.6]
\end{displaymath}

Consider the above controllers $C_{A}$ and $C_{B}$ and recall that these values correspond to commanded wheel velocities. It is clear simply from examination that $C_{A}$ is a controller that will perform very slowly in comparison to $C_{B}$. By taking advantage of this information, and other investigative tactics, it is possible to make some general distinctions about behaviors by examining their controllers without simulation.

We extend this principle to attempt to identify three undesirable swarm characteristics.

\begin{figure*}
     \centering
     \begin{subfigure}[b]{0.18\linewidth}
         \centering
         \includegraphics[width=\linewidth]{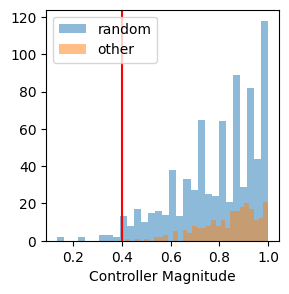}
         \caption{Element Magnitude}
         \label{fig:heuristic-element-magnitude}
     \end{subfigure}
     \hfill
     \begin{subfigure}[b]{0.18\linewidth}
         \centering
         \includegraphics[width=\linewidth]{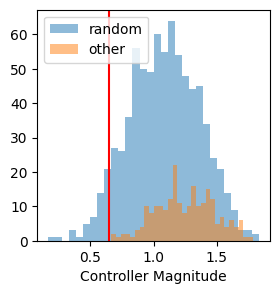}
         \caption{Controller Magnitude}
         \label{fig:heuristic-controller-magnitude}
     \end{subfigure}
     \hfill
     \begin{subfigure}[b]{0.18\linewidth}
         \centering
         \includegraphics[width=\linewidth]{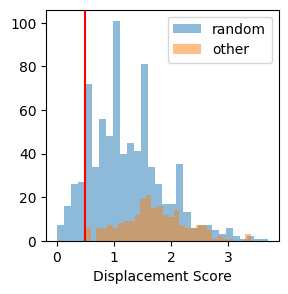}
         \caption{Displacement}
         \label{fig:heuristic-displacement}
     \end{subfigure}
     \hfill
     \begin{subfigure}[b]{0.18\linewidth}
         \centering
         \includegraphics[width=\linewidth]{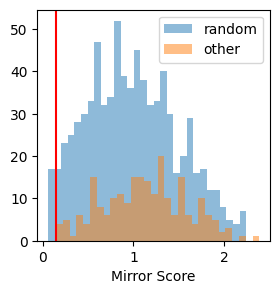}
         \caption{Mirror Distance}
         \label{fig:heuristic-mirror-distance}
     \end{subfigure}
     \hfill
     \begin{subfigure}[b]{0.18\linewidth}
         \centering
         \includegraphics[width=\linewidth]{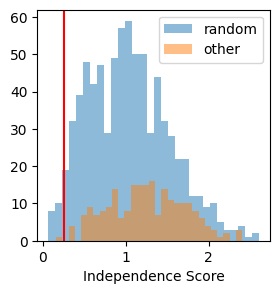}
         \caption{Neglectful Distance}
         \label{fig:heuristic-neglectful-distance}
     \end{subfigure}
     \hfill
     
    \caption{Scores of random (blue) and emergent (orange) behaviors on a set of 5 controller metrics. From left to right: (a) the maximum absolute value of the controller velocities, (b) the L2 norm of the controller, (c) The Displacement metric, representing the controller's spin, (d) the L2 distance from the controller to velocities that will perform the reverse behavior when it's sensors are on, and (e) the L2 distance from the controller to velocities that will neglect other agents. Red lines indicate threshold values for $\psi$, where scores to the left of the line reflect random behaviors only.}
    \label{fig:heuristic components}
\end{figure*}

\subsection{Slow Swarms} Agents that operate with very low velocities will not converge upon a meaningful collective behavior within the simulation horizon, $t$. We consider a controller to be slow if the element of $C$ with the greatest magnitude (L1 distance from 0.0) is less than some threshold $\psi_{1}$. 

\begin{equation}
    max(-min(C), max(C)) < \psi_1
    \label{eq: hueristic A}
\end{equation}

Similarly, controller $C$ is considered slow if the magnitude of $C$ falls below some threshold $\psi_{2}$.

\begin{equation}
    ||C||_{2} < \psi_2
    \label{eq: hueristic B}
\end{equation}

\subsection{Passive Swarms} Agents that fail to explore outside of their initial state will not converge upon meaningful behavior. Our Heuristic detects two instances of this. The first is where behaviors spin in place, resulting in low displacement in the environment. For a controller $C = [a, b, c, d]$, $C$ is considered passive, and low displacement, if 

\begin{equation}
    |a + b| + |c + d| < \psi_3
    \label{eq: hueristic C}
\end{equation}

A controller with low displacement will contain velocities that  reduce to "spinning in place". Consider Equations~\ref{eq: kinematics x} and \ref{eq: kinematics y} from our swarm kinematics. Even when wheels have high velocities, if $v_l = -v_r$, the resulting kinematics results in a large angular velocity but no change in x and y position. We seek behaviors that will explore locally, therefore, we prioritize controllers with $|a + b| + |c + d| < \psi_3$.

Similarly, if the behavior executed when the sensor is on is exactly opposite the behavior executed when the sensor does not detect anything, resulting in agents that go forward and then backwards in the same motion. We refer to this as a mirrored controller, where the velocites when the sensor is off are exactly opposite the velocities when the sensor is on. For a controller, $C = [a, b, c, d]$, and its mirror controller, $C' = [a, b, -a, -b]$, we consider the controller passive if 

 \begin{equation}
     dist([a, b, c, d], [a, b, -a, -b]) < \psi_4
    \label{eq: hueristic D}
\end{equation}

\subsection{Neglectful Swarms} An Agent neglects the swarm if sensing another agent does not trigger a significant change in its controller. Neglectful agents fail to uncover meaningful behavior. We consider an agent to be neglectful if for a controller $C = [a, b, c, d]$, the distance to the neglectful controller is below some threshold, 

\begin{equation}
     dist([a,b, c, d], [a, b, a, b]) < \psi_5
    \label{eq: hueristic E}
\end{equation}

Combined, these five metrics are used to filter out undesirable behaviors based on controller intuition alone. We select values for our $\psi$ thresholds by examining the distribution of labeled behaviors when we apply our 5 heuristic criteria on them, as shown in Fig.~\ref{fig:heuristic components}.

\begin{table}[h]
\begin{tabular}{|c|c|c|c|c|}
\hline
$\psi_1$ & $\psi_2$ & $\psi_3$ & $\psi_4$ & $\psi_5$ \\ \hline
0.4 & 0.65 & 0.5 & 0.2 & 0.3 \\ \hline
\end{tabular}
\end{table}

\begin{figure}[t]
     \centering
     \includegraphics[width=0.8\linewidth]{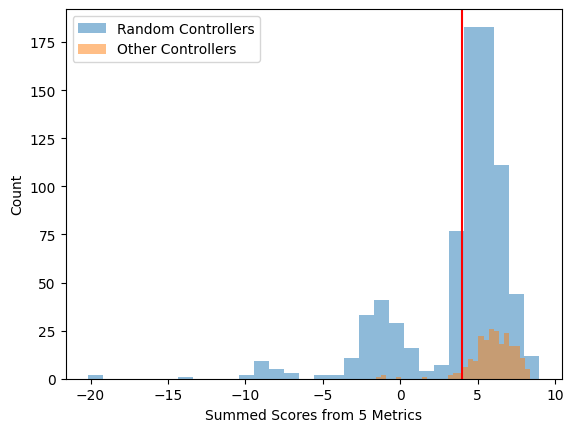}
     \caption{Distribution of heuristic scores on a labeled set of 1000 swarm behaviors. The red line indicates a threshold such that $>95\%$ of the controllers that score below this line are  random.}
    \label{fig:combined heuristics}
\end{figure}

Equations \ref{eq: hueristic A} through \ref{eq: hueristic E} are evaluated compared to the thresholds shown in the table above. If the calculated metric is less than the threshold, a penalty of -5 is applied to the score, otherwise the left hand side of all equations is summed together to a controller's heuristic score. We show the associated histogram and cutoff threshold in Figure \ref{fig:combined heuristics}.

We find that out of an initial sample of 1000 controllers, 777 of them were classified as random controllers. Our heuristic correctly identifies 230 (29.6\%) as random, while only removing 12 emergent behaviors from consideration. In the case of our robot models, these 12 behavior classes were still represented in the unfiltered group, but in some cases the threshold values will need to be tuned to guarantee that only random behaviors are being filtered out.

It is worth noting that our goal in filtering is not to completely eliminate the random controllers, but instead to confidently decrease the size of our search space without the removal of entire behavior classes. Under the assumption that slow, passive, and neglectful swarms are not of interest to the human, these heuristics provide a good starting point for future reduction of the search space.

\begin{figure*}
     \centering
     \begin{subfigure}[b]{0.33\linewidth}
         \centering
         \includegraphics[width=\linewidth]{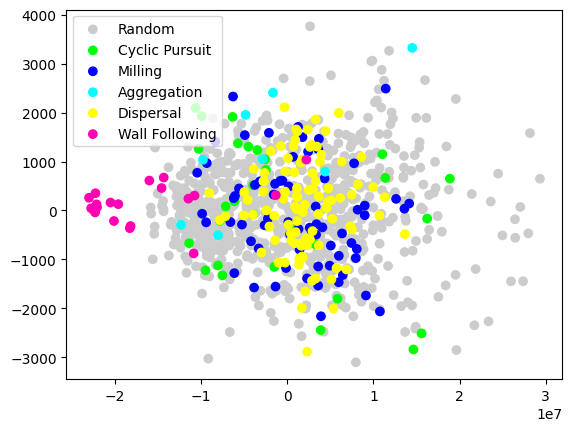}
         \caption{Random initialization}
         \label{fig:vis-embedding-random}
     \end{subfigure}
     \hfill
     \begin{subfigure}[b]{0.33\linewidth}
         \centering
         \includegraphics[width=\linewidth]{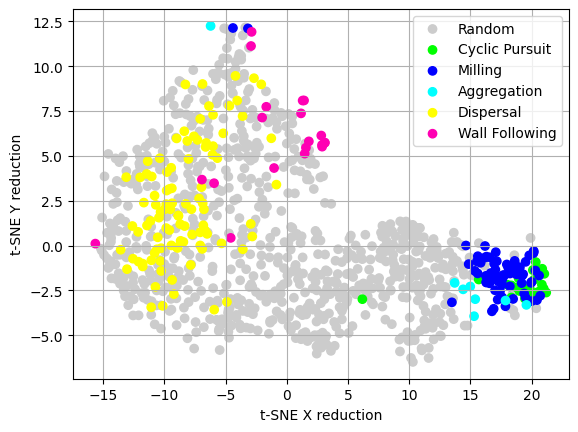}
         \caption{Hand-Crafted Metrics}
         \label{fig:vis-embedding-hand-crafted}
     \end{subfigure}
     \hfill
     \begin{subfigure}[b]{0.33\linewidth}
         \centering
         \includegraphics[width=\linewidth]{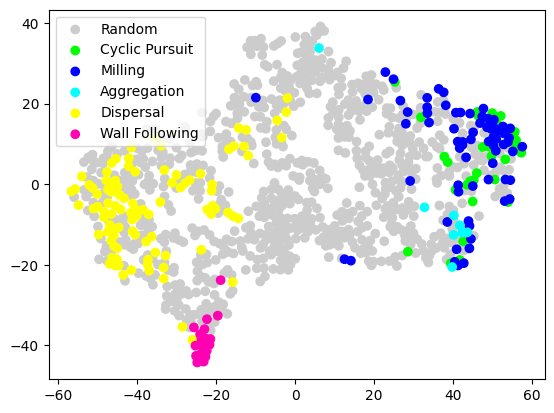}
         \caption{HIL Latent Learning (Ours)}
         \label{fig:vis-latent-learning}
     \end{subfigure}
    \caption{t-SNE visualizations of behavior similarity in latent spaces. From left to right: (a) Our randomly-initialized network with no training, (b) Brown et al's~\cite{brown2018discovery} hand-crafted behavior mapping, and (c) our learned latent space trained with contrastive learning and human-in-the-loop queries.}
    \label{fig:embedding visualizations}
\end{figure*}

\begin{figure}[t]
     \centering
     \includegraphics[width=\linewidth]{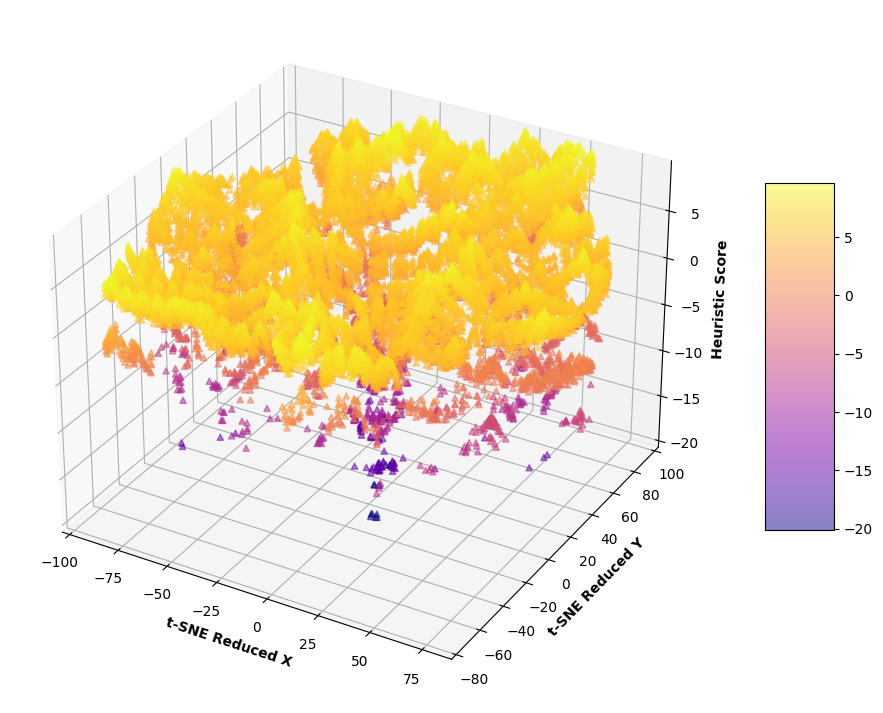}
     \caption{Heuristic scores for the discretized single sensor controller space. Values below $z=4.0$ are filtered from evolution and network training.}
    \label{fig:3d-heuristics}
\end{figure}

We also include a visualization of heuristic scores the controller space when heuristic scores are applied. Figure \ref{fig:3d-heuristics} shows the entirety of our discretized single-sensor controller space, where the x and y axis are the t-SNE embeddings into two dimensions and the z axis is our heuristic score. Our heuristic filters 43,251 (22.2\%) of the 194,481 controllers from the discretized controller space.

\section{Latent Space Visualizations} \label{Appendix: Latent Visualizations}
In conjunction with the accuracy calculations performed on our models (Table \ref{table:Embedding Space Accuracy}), we present visualizations of the evaluated latent spaces using a t-SNE embedding~\cite{van2008visualizing} to reduce the behavior representations from $\mathbb{R}^5$ to $\mathbb{R}^2$.

We take a set of 1000 labeled behaviors from the single-sensor space and determine the embedding of each behavior for our randomly initialized deep network, Brown et al's hand-crafted vector and our latent space with HIL and Filtering. Figure \ref{fig:embedding visualizations} shows the clusters of 6 behaviors in the 2D plane.

Inspection of this visualizations supports the results we obtained in our paper, showing that a randomly initialized network will poorly embed behaviors into the space with respect to feature differences (Figure \ref{fig:vis-embedding-random}). In a hand-crafted behavior space (Figure \ref{fig:vis-embedding-hand-crafted}), behavior distinction is more clear, but the Wall-Following behavior in particular is poorly embedded, and distinctions between milling, cyclic pursuit and aggregation are difficult to distinguish. In our Latent Learning approach (Figure \ref{fig:vis-latent-learning}), we show that dispersal and wall following are clearly distinct in their embedding and our approach does better than hand-crafted on distinctions between cyclic pursuit and aggregation, but still has difficulty distinguishing the milling behavior.

% \subsection{Euclidean Proximity (L2 Accuracy)}

% \subsection{Triplet Loss Minimization (Margin Accuracy)}

% \section{Latent Space Visualizations}
% Using t-SNE dimensionality reduction, we offer 

% \section{Human Querying}

% \section{Evolutionary Parameter Optimization}

\section{Example Controllers}
We provide controllers for each behavior uncovered by our experiments. These controllers are designed for the parameters, dimensions, and assumptions presented about our simulator and agents in Appendix \ref{Appendix: Swarm Kinematics}.

Milling: $[0.6, 1.0, 0.4, 0.5]$,

Cyclic Pursuit: $[-0.7, 0.3, 1.0, 1.0]$,

Aggregation: $[-0.7, -1.0, 1.0, -1.0]$,

Dispersal: $[0.2, 0.7, -0.5, -0.1]$,

Wall Following: $[1.0, 0.9, 1.0, 1.0]$,

Random: $[-0.8, -0.7, 0.2, -0.5]$,

Nested Cycle: $[0.8, 0.5, 0.6, -0.5, -0.5, 0.0, -0.2, 0.5, -\frac{\pi}{3}]$,

Concave Cycle: $[-0.4, 0.8, 0.9, -0.1, 0.6, 1.0, 0.4, 0.0, \frac{\pi}{6}]$

We include videos of the simulation of each behavior on our website.

\end{document}